\documentclass[aps,prb,twocolumn,showpacs]{revtex4} 

\usepackage{amsmath,amssymb,mathrsfs}
\usepackage{latexsym}
\usepackage{graphicx} 
\usepackage{epstopdf}
\usepackage{graphicx,epstopdf,color}
\usepackage{amsfonts}
\usepackage{dsfont} 
\usepackage{bm}

\begin{document}
\renewcommand{\arraystretch}{1.7}
\renewcommand{\tabcolsep}{0.3cm}
\newcommand{\1}{\hspace*{-1pt}}
\newcommand{\2}{\hspace*{-2pt}}
\newcommand{\3}{\hspace*{-3pt}}

\title{Gate defined wires in HgTe quantum wells: from Majorana fermions to spintronics}

\author{Johannes Reuther}
\affiliation{Department of Physics, California Institute of Technology, Pasadena, CA 91125 USA}
\author{Jason Alicea}
\affiliation{Department of Physics, California Institute of Technology, Pasadena, CA 91125 USA}
\author{Amir Yacoby}
\affiliation{Department of Physics, Harvard University, Cambridge, Massachusetts 02138 USA}

 \pagestyle{plain}

\begin{abstract}
We introduce a promising new platform for Majorana zero-modes and various spintronics applications based on gate-defined wires in HgTe quantum wells.  Due to the Dirac-like band structure for HgTe the physics of such systems differs markedly from that of conventional quantum wires.  Most strikingly, we show that the subband parameters for gate-defined HgTe wires exhibit exquisite tunability: modest gate voltage variation allows one to modulate the Rashba spin-orbit energies from zero up to $\sim30$K, and the effective g-factors from zero up to giant values exceeding $600$. The large achievable spin-orbit coupling and g-factors together allow one to access Majorana modes in this setting at exceptionally low magnetic fields while maintaining robustness against disorder.  As an additional benefit, gate-defined wires (in HgTe or other settings) should greatly facilitate the fabrication of networks for refined transport experiments used to detect Majoranas, as well as the realization of non-Abelian statistics and quantum information devices. 
\end{abstract}


\maketitle

\section{Introduction}

The ability to efficiently manipulate electron spins with electric and magnetic fields underlies a wide variety of solid-state applications.\cite{zutic04}  Prominent classic examples include giant magnetoresistance,\cite{baibich88,GMR} spin qubits,\cite{SpinQubits,koehl11,pla12} and spin transistors.\cite{datta90,jansen03,zutic04}  Recent proposals for stabilizing Majorana zero-modes in topological insulator\cite{fu08,fu09,cook11} and semiconductor\cite{lutchyn10,oreg10,sau10,alicea10} architectures, while not usually viewed from a spintronics lens, similarly rely crucially on spin manipulation.  In essence these approaches utilize spin-orbit coupling and Zeeman fields to effectively convert an ordinary $s$-wave superconductor into a `spinless' topological superconductor supporting Majorana zero-modes (for recent reviews, see Refs.~\onlinecite{beenakker11,alicea12,leijnse12,stanescu13}).  Intense experimental efforts, driven partly by potential quantum computing applications\cite{kitaev03,nayak08}, have already delivered possible Majorana signatures.\cite{mourik12,das12,rokhinson,deng12,finck12}  

\begin{figure*}[t]
\centering
\includegraphics[scale=0.5]{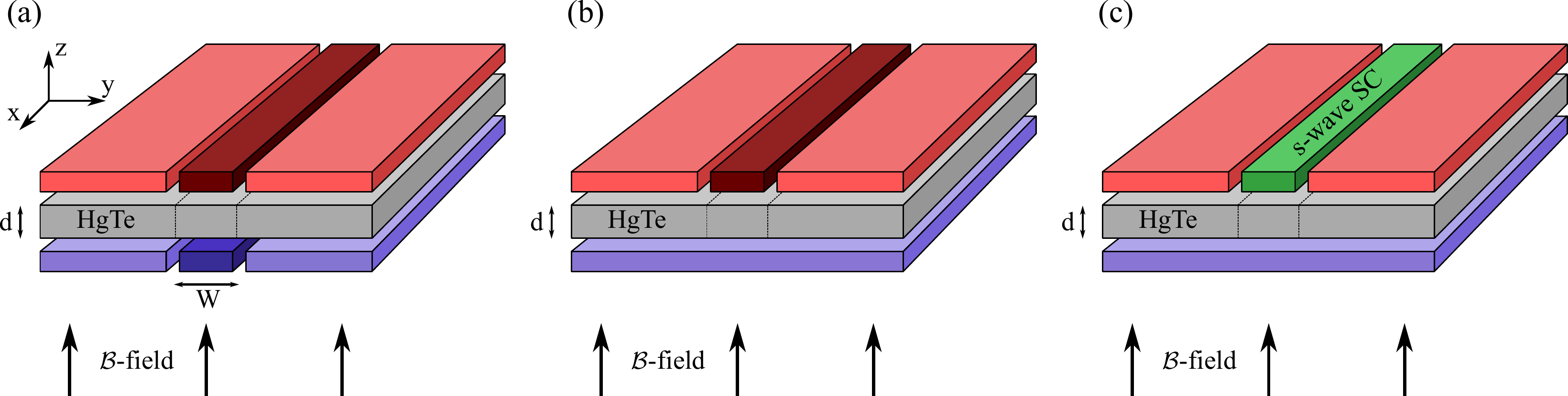}
\caption{Schematic experimental setups. (a) A 2D HgTe quantum well of thickness $d$ sandwiched by top and bottom gates and impinged by a magnetic field.  The outer regions are tuned into an insulating state while the central gates define a quantum wire.  Due to HgTe's unusual band structure, properties of the quantum wire (such as Rashba coupling and Zeeman splitting generated by a perpendicular magnetic field) are exceptionally tunable.  (b) Simplified experimental setup featuring a single bottom gap.  In this more experimentally feasible device, parameters for the wire exhibit essentially the same level of tunability as discussed in Sec.\ \ref{subsec:eso}.  (c) Replacing the central top gate by a proximity coupled $s$-wave superconductor allows the gate-defined quantum wire to host Majorana zero-modes.  This application is especially attractive given the large spin-orbit energies ($\sim30K$) and giant effective $g$-factors (which can reach $\sim600$) that are achievable.  }
\label{fig:setup_hgte}
\end{figure*}

For many such spin-based applications, materials exhibiting easily tunable spin-orbit coupling and $g$-factors are highly desirable.  In this paper we employ complementary analytical and numerical methods to demonstrate that gate-defined wires in HgTe quantum wells (see the geometries in Fig.\ \ref{fig:setup_hgte}) satisfy both criteria.  By itself this observation is unremarkable; for instance, Rashba coupling in semiconductors is well-known to be gate-tunable,\cite{nitta97} while $g$-factors can be modified through various means including electric fields and strain.\cite{snelling91,malinowski00,Maier}  Rather, the special feature of the HgTe wires we study---which stems largely from the unusual Dirac-like band structure exhibited by the quantum well---lies in the extraordinary degree to which these parameters can be controllably varied under realistic conditions.  

As in any semiconductor, gate voltages can induce moderate changes in Rashba coupling for the two-dimensional HgTe quantum well hosting the wire.  We show, however, that the \emph{effective} Rashba parameters for quasi-one-dimensional confined subbands vary much more dramatically and in an oscillatory fashion, similar to Refs.\ \onlinecite{mireles01,governalea04,knobbe05,zhang09}.  Relatively modest gate voltages can consequently alter the characteristic spin-orbit energies for the wire from zero to appreciable values of $\sim 30$K (for comparison typical spin-orbit energies for electron-doped wires such as InAs or InSb are $\sim 1$K\cite{nadjperge12}).  More surprising is the behavior of the effective $g$-factors for confined subbands, which in contrast to typical wires are by far dominated by orbital contributions from the magnetic field (at least when directed normal to the well).  These $g$-factors similarly undergo gate-induced oscillations and can be driven from zero to enormous values exceeding $600$ due to orbital enhancement.  In both cases the remarkable oscillatory dependence originates from non-perturbative modifications of confined wavefunctions in response to gating.  

Because of this exquisite tunability, gate-defined HgTe wires are prime candidates for spintronics and related applications.  Here we focus on one particularly enticing example---the pursuit of Majorana modes for topological quantum information processing.  (Note that edge states of HgTe in the two-dimensional topological insulator phase can also host Majoranas.\cite{fu09,bernevig06,koenig07,koenig08,nowack12,HgTeCurrents}  The physics we discuss here is unrelated to these edge states, but is instead close in spirit to the semiconductor wire proposals from Refs.\ \onlinecite{lutchyn10,oreg10}.)  We show that when a good proximity effect with an $s$-wave superconductor is generated, the giant $g$-factors allow for exceptionally weak fields---a few tens of mT---to drive the wire into a topological superconductor with Majorana zero-modes.  The strong spin-orbit coupling for the HgTe wire (compared to typical electron-doped wires) further allows this topological state to possess a relatively large gap that exhibits enhanced immunity against disorder.\cite{potter11}  Apart from these virtues we expect that gate-defined wires offer another important longer-term advantage as well.  Namely, synthesizing arbitrary wire networks merely requires patterning of additional gates on the quantum well.  These can serve the dual purpose of enabling refined multi-terminal transport detection of a topological phase transition and Majorana zero-modes, along with braiding of Majoranas to harness their non-Abelian statistics.\cite{alicea11,clarke,vanheck12,halperin12}  Such benefits provide strong motivation for pursuing Majorana physics in gate-defined wires in HgTe or related platforms. 

The remainder of this paper is structured as follows.  Section \ref{sec:system} explores the physical properties of the HgTe wires.  We then turn in Sec.\ \ref{sec:pairing} to the application of Majorana zero-modes in this setting, treating both the clean and disordered cases within a simplified framework.  Section \ref{sec:conclusion} summarizes our main results and discusses future directions in greater detail.  Finally, three appendices contain additional calculations that further support the claims in this paper.

\section{Characterization of gate-defined wires}\label{sec:system}

In this section we perform a detailed characterization of gate-defined HgTe wires.  Section \ref{subsec:band} develops an analytical description of the system, starting from the Hamiltonian for a 2D HgTe quantum well and then systematically including the effects of a confinement potential, Rashba coupling, and applied magnetic fields.  More accurate numerical simulations are explored in Sec.\ \ref{subsec:num}.  Our objective below is to demonstrate that such wires exhibit large and exceptionally tunable Rashba spin-orbit coupling and $g$-factors as claimed in the introduction, rendering them promising for applications that will be briefly discussed in Secs.\ \ref{sec:pairing} and \ref{sec:conclusion}.  

\subsection{Analytic treatment of the confinement problem}
\label{subsec:band}

We begin by reviewing the well-studied physics of uniform quasi-2D HgTe quantum wells, following conventions used in Refs.\ \onlinecite{bernevig06} and \onlinecite{koenig08}.  Excitations in the four bands closest to the Fermi level can be described with a spinor $\Psi({\bf r})^T=[\psi_{E+}({\bf r}),\psi_{H+}({\bf r}),\psi_{E-}({\bf r}),\psi_{H-}({\bf r})]$.  Here $\psi_{E,s}({\bf r})$ and $\psi_{H,s}({\bf r})$ annihilate states with opposite parity at position ${\bf r} = (x,y)$ in the quantum well plane.  Under time-reversal $\mathcal{T}$ these operators transform as $\psi_{E/H,s}\rightarrow s\psi_{E/H,-s}$.  The standard model Hamiltonian for the 2D quantum well reads\cite{bernevig06,koenig08}
\begin{equation}
  H_{2D}=\int d^2{\bf r}\Psi^{\dagger} \Bigg(\begin{array}{cc}
h & 0 \\
0              & h^*\\
\end{array}\Bigg)\Psi
  \label{general_ham}
\end{equation}
with
\begin{align}
h&=[-\mu_{2D}+D (\partial_x^2+\partial_y^2)]I+{\bf d}\cdot {\bm \sigma}
 \nonumber \\
{\bf d}&=[-iA\partial_x,iA\partial_y,M+B(\partial_x^2+\partial_y^2)]. 
\label{ham_hgte2} 
\end{align}
In the above equations $I$ is a $2\times 2$ identity matrix, $\bm \sigma$ denotes a vector of Pauli matrices, $\mu_{2D}$ represents the chemical potential, and $A$, $B$, $D$, and $M$ are materials parameters dependent on the quantum-well thickness $d$.  

For the remainder of this subsection we focus on long-wavelength physics where it suffices to set $B = D = 0$, as doing so greatly facilitates analytic treatment of the problem.  Our numerics in Sec.\ \ref{subsec:num} restore these terms to experimentally relevant values and confirm that they do not change the physics qualitatively.  With this simplification Eq.\ (\ref{general_ham}) describes massive Dirac fermions with band energies 
\begin{equation}
  E_\pm({\bf k}) = -\mu_{2D} \pm \sqrt{M^2 + (A k)^2}.
  \label{Dirac_spectrum}
\end{equation}
The gap $2M$ for HgTe wells is quite small---typically on the order of 0.01eV.\cite{bernevig06,koenig08}  As emphasized in the introduction the Dirac structure together with this small mass cause the properties of gate-induced confined states in the bulk of HgTe to differ dramatically from those in conventional semiconductors such as GaAs or InAs.  We comment further on such distinctions below.

\begin{table}[t]
\centering
\begin{tabular}{lr}
\hline
$M$[eV] & -0.01\\
$A$[eV\AA] & 3.65\\
$B$[eV\AA$^2$] & -68.6\\
$D$[eV\AA$^2$] & -51.2\\
$R/(eE_z)$[\AA$^2$] &  -1560\\
$g_E$ & 22.7\\
$g_H$ & -1.21\\
\hline
\end{tabular}
\caption{Parameters for a 2D HgTe quantum well with thickness $d=70$\AA.\cite{koenig08}  At such a thickness the system happens to realize a topological insulator, though this property is inconsequential for the formation of gate-defined wires in the bulk of the quantum well.}
\label{table}
\end{table}

Suppose that one now couples the quantum well to a set of top and bottom gates as shown in Fig.\ \ref{fig:setup_hgte}(a).  [The essential physics exhibited by this system can also be captured in the simpler experimental setup of Fig.\ \ref{fig:setup_hgte}(b), which contains only a single bottom gate; we discuss this further in Sec.\ \ref{subsec:eso}.]  These gates allow one to separately tune the global chemical potential, the perpendicular electric field in each region,\cite{nitta97} and the relative potential between inner and outer regions which will define a quantum wire of width $W$. Throughout this paper we assume that the voltages on the left and right pairs of gates are tuned identically to fully deplete carriers from the outer regions of HgTe.  For now we also assume that each pair of top and bottom gates is adjusted symmetrically so that structural inversion ($z\rightarrow-z$) symmetry is present.  (This restriction will be relaxed below when we discuss Rashba coupling.)  If $\mathcal{V}(y)$ denotes the confinement potential defining the wire, then under these conditions the Hamiltonian becomes $H = H_{2D} + H_{\rm conf}$ where
\begin{equation}
  H_{\rm conf} = \int d{\bf r} \mathcal{V}(y) \Psi^\dagger \Psi.
  \label{Hconf}
\end{equation}

\begin{figure}[t]
\centering
\includegraphics[scale=0.75]{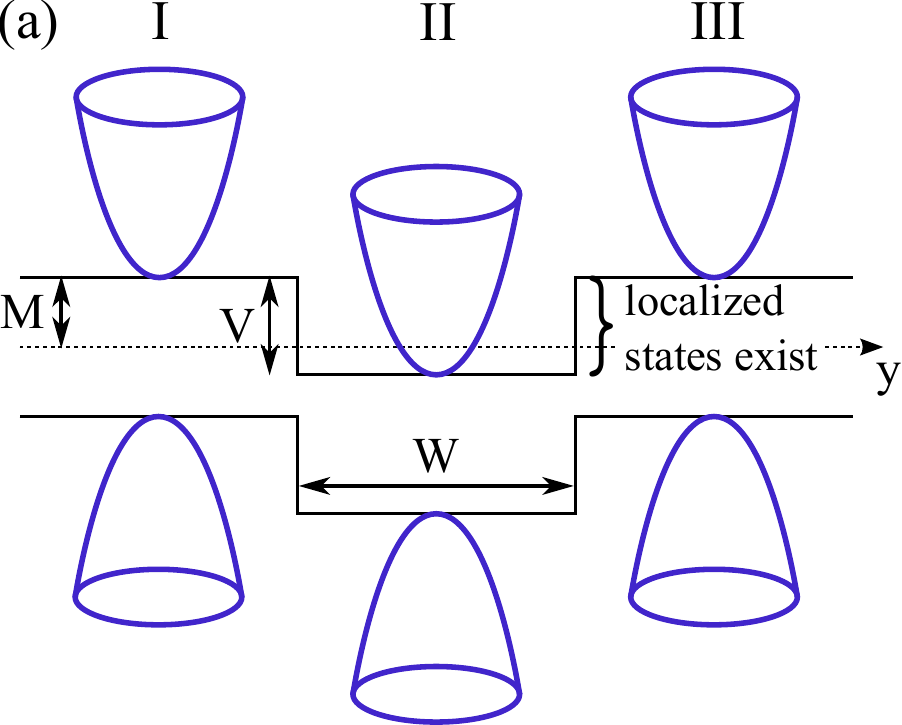}\\
\includegraphics[scale=0.58]{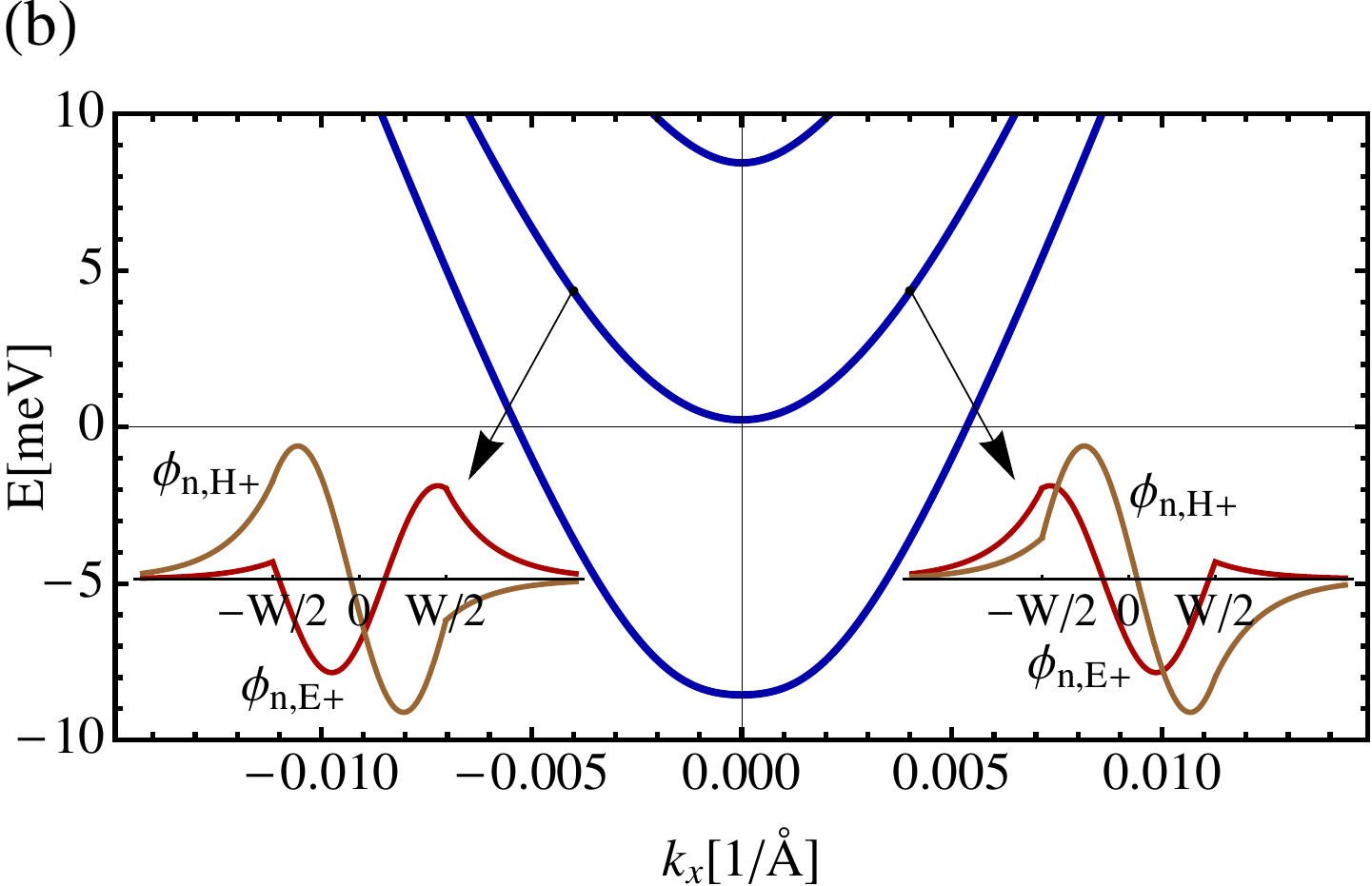}
\caption{(a) Schematic illustration of the gate-defined quantum wire. (b) Main plot: Band structure for a $W = 800$\AA~gate-defined quantum wire residing in a $d = 70$\AA~HgTe quantum well.  The materials parameters used appear in Table \ref{table}, while the depth of the confining potential is $V = 0.025$eV.  Note that each band is doubly degenerate.  Insets: Illustration of the wavefunction components $\phi_{n,E/H+}$ in the $n = 2$ band.}
\label{fig:bands}
\end{figure}

Here we model the confinement with $\mathcal{V}(y)=-V\Theta(W/2-|y|)$, though a more realistic smooth confinement potential will be treated numerically later in Sec.\ \ref{subsec:num}.  As illustrated in Fig.\ \ref{fig:bands}(a) states localized along $y$ can exist in an energy window $\Delta E = {\rm min}(|M|,|V|)$.  One can derive an effective 1D Hamiltonian for these confined subbands---which we label by an index $n = 1,2,\ldots$---by projecting the 2D quantum well Hamiltonian using
\begin{eqnarray}
  \Psi({\bf r}) &\rightarrow& \sum_{n} \int_{k_x}e^{i k_x x}\bigg\{ 
  \left[\begin{array}{c}
	\Phi_{n+}(k_x,y) \\
	0
  \end{array}\right]\psi_{n+}(k_x) 
  \nonumber \\
  &+& \left[\begin{array}{c}
	0 \\
	\Phi_{n-}(k_x,y)
  \end{array}\right]\psi_{n-}(k_x)\bigg\}.
  \label{projection}
\end{eqnarray}  
The operators $\psi_{n\pm}(k_x)$ above correspond to Kramer's pairs and annihilate states in the gate-defined wire with momentum $k_x$ in band $n$.  One can obtain the two-component wavefunctions $\Phi_{n\pm}=(\phi_{n,E\pm},\phi_{n,H\pm})^T$ and associated band energies in the standard way by solving the Hamiltonian separately in the three regions of Fig.\ \ref{fig:bands}(a) and then matching boundary conditions (see Appendix \ref{appendix1} for details).  Below we simply highlight some salient features of the problem.  

First, unlike for a conventional parabolic 2D dispersion, the $x$- and $y$-directions cannot be treated independently---hence the wavefunctions $\Phi_{n,\pm}(k_x,y)$ depend on $k_x$.  We define overall phases such that $\Phi_{n,s}$ are purely real (which is always possible due to the form of $H$); moreover, these functions satisfy 
\begin{eqnarray}
  \Phi_{n+}(k_x,y)&=&(-1)^{n} \sigma^z\Phi_{n+}(-k_x,-y) 
  \nonumber \\
  \Phi_{n+}(k_x,y) &=& \Phi_{n-}(-k_x,y).
  \label{PhiProperties}
\end{eqnarray}  
Note that except at $k_x = 0$ $\Phi_{n,s}(k_x,y)$ does not have well-defined parity under $y\rightarrow -y$.  It follows from the properties above that inversion sends $\psi_{n,s}(k_x) \rightarrow (-1)^{n} \psi_{n,s}(-k_x)$ while under time reversal $\psi_{n,s}(k_x) \rightarrow s\psi_{n,-s}(-k_x)$.

Generally, increasing $W$ reduces the energy difference between the confined bands, which scales like $\sim \pi A/W$ for $|V|\gg|M|$. Increasing the depth $V$ of the confining potential shifts these bands down in energy, allowing new confined states to emerge from the upper half of the Dirac cone.  For $V \gtrsim 2M$ the lowest-energy confined bands begin to merge with the lower half of the Dirac cone; states in these bands remain confined at `large' $k_x$ but are extended at `small' $k_x$ due to hybridization with bulk states.  This feature will be important in our numerics discussed in Sec.\ \ref{subsec:num}.  Within our analytical treatment, however, we avoid this complication for simplicity.

Figure \ref{fig:bands}(b) illustrates the confined band energies and wavefunctions for a $d = 70$\AA~quantum well hosting a gate-defined wire of width $W = 800$\AA~and potential depth $V = 0.025$eV.  To generate these curves we employed parameters from Table \ref{table}, which lists various quantities relevant for 70\AA-thick HgTe sheets\cite{koenig08}.  These values will in fact frequently be adopted in our simulations below since for this thickness quantitative estimates are known for most parameters of interest to us here.  As an important aside, we note that for the quoted ratio of $M/B$ in the table HgTe resides in the topological insulator phase.\cite{koenig08}  We stress, however, that whether the topological or trivial state appears has little bearing on the existence of confined gapless 1D states that we seek to generate in the system's \emph{bulk}.  (Of course changing $d$ to enter the trivial phase modifies the parameters in Table \ref{table} and therefore has a quantitative effect on properties of the confined states.  But the important point is that there is no sharp distinction in the two cases insofar as these levels is concerned.)

Upon expanding the confined band energies to second order in $k_x$, one obtains a simple effective 1D quantum wire Hamiltonian which serves as the starting point for our analysis in this paper:
\begin{equation}
  H_{1D}=\int_{k_x}\sum_{n}\psi_{n}^{\dagger}\left(\epsilon_n -\mu+ \frac{\hbar^2k_x^2}{2m_n}\right)\psi_{n},
  \label{H1D}
\end{equation}
where $\epsilon_n$ and $m_n$ respectively denote the subband energy and effective mass for band $n$, $\mu$ is the effective chemical potential, and the sum over pseudospin $s = \pm$ is left implicit.  A rough estimate for the effective masses of bands far from extended states can be obtained by setting $k_y = n\pi/W$ in Eq.\ (\ref{Dirac_spectrum}) and then expanding the Dirac spectrum to order $k_x^2$; this yields $m_n\sim (|M|/A^2)\sqrt{1+[n\pi A/(M W)]^2}$.  Inserting parameters from Table \ref{table} and assuming $W = 800$\AA, one finds that $|M|/A^2 \approx 0.006m_e$, where $m_e$ is the bare electron mass, while $[\pi A/(M W)]^2\approx 2$.  For these parameters the effective mass therefore increases appreciably with $n$.  As an illustration, $m_n$ is $\sim0.01m_e$ for $n = 1$ (which is comparable to the effective mass for electron-doped InSb) but is enhanced to $0.04m_e$ for $n = 5$ (which is close to the effective mass for electron-doped GaSb).  Gate-defined wires in systems with conventional parabolic bands, by contrast, exhibit masses that to a first approximation are independent of the band index.  

Because of the symmetries imposed so far, at a given momentum each band is doubly degenerate.  For applications---\emph{e.g.}, the pursuit of Majorana fermions---it is highly desirable to lift this degeneracy via perturbations such as Rashba spin-orbit coupling and applied magnetic fields.\cite{lutchyn10,oreg10} We turn now to incorporating these ingredients into our effective 1D Hamiltonian.

\subsubsection{Effective Rashba coupling}
\label{subsec:eso}

Suppose now that the top and bottom gates in Fig.\ \ref{fig:setup_hgte}(a) are adjusted asymmetrically.  Such asymmetric gating generates a perpendicular electric field $E_z$, resulting in a voltage drop $U=E_z d$ across the quantum well width $d$.  The loss of structural inversion symmetry leads to Rashba spin-orbit coupling, which for simplicity we assume is induced uniformly throughout the 2D quantum well (this is by no means essential).  Our objective here is to explore the \emph{effective} Rashba coupling felt by the confined electrons in our gate-defined wire.  We will begin with the regime where the electric field $E_z$ is weak (in a sense to be quantified below) so that one can extract this effective coupling within first-order perturbation theory.  This perturbative analysis provides rough order-of-magnitude estimates for the achievable spin-orbit energies characterizing the wire.  We emphasize, however, that for `large' $E_z$ modifications of the confined wavefunctions produce striking \emph{non-perturbative} effects which underlie our main findings in this paper.  An initial discussion of non-perturbative effects is provided below; additional results appear in Sec.\ \ref{subsec:num} and Appendix \ref{appendix3}.  

Our perturbative analysis begins with the known Rashba Hamiltonian for the 2D HgTe quantum well,\cite{rothe10}
\begin{equation}
  H_R = \int d^2{\bf r}\Psi^\dagger
  \left(\begin{array}{cccc}
    0                                         &0 &R(-\partial_x+i\partial_y) &0\\
    0                                         &0 &0                                          &0\\
    R(\partial_x+i\partial_y) &0 &0                                          &0\\
    0                                         &0 &0                                          &0
  \end{array}\right)\Psi. 
  \label{ham_rashba}
\end{equation} 
Here $R=Fe U$, with $e>0$ the magnitude of the electron charge and $F$ a material (and geometry) dependent parameter.  Upon projecting $H_R$ onto the confined bands using Eq. (\ref{projection}), one obtains an effective Rashba Hamiltonian for the gate-defined wire,
\begin{equation}
  H_R \rightarrow H_{1D,R} = \int_{k_x} \sum_{n n'}[-ir_{n n'}(k_x)\psi_{n+}^\dagger\psi_{n'-}+H.c.],
\end{equation}
which contains both intraband \emph{and} interband couplings of strength
\begin{equation}
  r_{n n'}(k_x) = R \int dy\phi_{n,E+}(k_x,y)(k_x-\partial_y)\phi_{n',E-}(k_x,y).
  \label{r}
\end{equation}
Time-reversal symmetry requires $r_{n n'}(k_x) = -r_{n'n}(-k_x)$ while properties of the wavefunctions in Eqs.\ (\ref{PhiProperties}) dictate that $r_{nn'}(-k_x) = (-1)^{n+n'+1}r_{nn'}(k_x)$.   Thus intraband Rashba couplings must be odd in $k_x$, as are interband couplings that mix bands with $n$ and $n'$ differing by an even integer; all other interband couplings are even in $k_x$.  For the moment we will assume that the electric field $E_z$ is sufficiently weak that all $r_{n n'}$ with $n\neq n'$ are small on the scale of the confined subband separation and can hence be ignored.  Continuing to focus on long-wavelength, low-energy physics, we expand the remaining intraband couplings as $r_{n n}(k_x) \approx \hbar \alpha_{n} k_x$ and neglect terms of order $k_x^3$ and higher.  Within these approximations, in the presence of Rashba coupling the 1D wire Hamiltonian in Eq.\ (\ref{H1D}) becomes
\begin{eqnarray}
  H_{1D}\rightarrow\int_{k_x}\sum_{n}\psi_{n}^{\dagger}\left(\epsilon_n -\mu + \frac{\hbar^2 k_x^2}{2m_n} + \hbar \alpha_n k_x \sigma^y \right)\psi_{n}. 
  \label{H1D2}
\end{eqnarray}

The Rashba coefficients $\alpha_n$ follow from Eq.\ (\ref{r}) and take the form $\alpha_n = c_n R/\hbar$, where $c_n$ are generically order-one dimensionless constants.   Together with the effective masses, these parameters define a characteristic spin-orbit energy for band $n$ via $E_{SO,n} = \frac{1}{2}m_n \alpha_n^2$.  Recalling that $R = F e U$ one can express this energy scale in terms of the voltage drop $U$ across the quantum well as
\begin{equation}
  E_{SO,n} = \frac{1}{2} m_n\left(\frac{c_n F e U}{\hbar}\right)^2.
  \label{E_SO_perturbative}
\end{equation}
To obtain rough numerical estimates consider a $d = 70$\AA~quantum well for which $F =\frac{R}{e E_z d}\approx -22.3$\AA~(using $E_z = U/d$ and $R/eE_z$ as given in Table \ref{table}).  For subbands with effective mass $m_n \sim 0.01 m_e$, a voltage drop $U \sim 0.05$V then yields a characteristic spin-orbit energy of $E_{SO,n} \sim 10$K.  Such scales reflect a roughly order-of-magnitude enhancement compared with spin-orbit energies in electron-doped InAs or InSb wires.\cite{nadjperge12,mourik12} 

Let us now quantify the range of $U$ over which the perturbative analysis above holds.  The physics is more universal for high subbands whose minimum is far in energy from the extended bulk states, so we focus on such cases for simplicity.  By inspecting Eq.\ (\ref{r}) one sees that interband couplings that mix adjacent subbands scale like $r_{n,n+1}(k_x = 0) \sim R/\lambda_y$, where $\lambda_y$ is the characteristic wavelength along $y$.  Roughly, $\lambda_y$ corresponds to the Fermi wavelength for electrons in region II of Fig.\ \ref{fig:bands}(a) so that $r_{n,n+1}(k_x = 0) \sim R V/A = Fe U V/A$ for large $n$.  Since the subband spacing scales like $\sim \pi A/W$, interband mixing is unimportant for voltage drops $U$ satisfying
\begin{equation}
  |U| \lesssim \frac{\pi A^2}{e|FV|W}~~~~({\rm perturbative ~ regime}).
  \label{perturbative_regime}
\end{equation}

For larger voltage drops interband mixing---not only with other confined bands but also typically with extended states since $M$ is rather small---becomes important.  The result is a dramatic reshaping of the confined wavefunctions by the perpendicular electric field, which has surprising and potentially useful consequences.  Specifically, upon increasing $U$ away from the perturbative regime, the effective Rashba energy characterizing a given confined band does not monotonically increase as one might naively expect, but instead undergoes striking \emph{oscillations}.  

\begin{figure*}[t]
\centering
\includegraphics[scale=0.85]{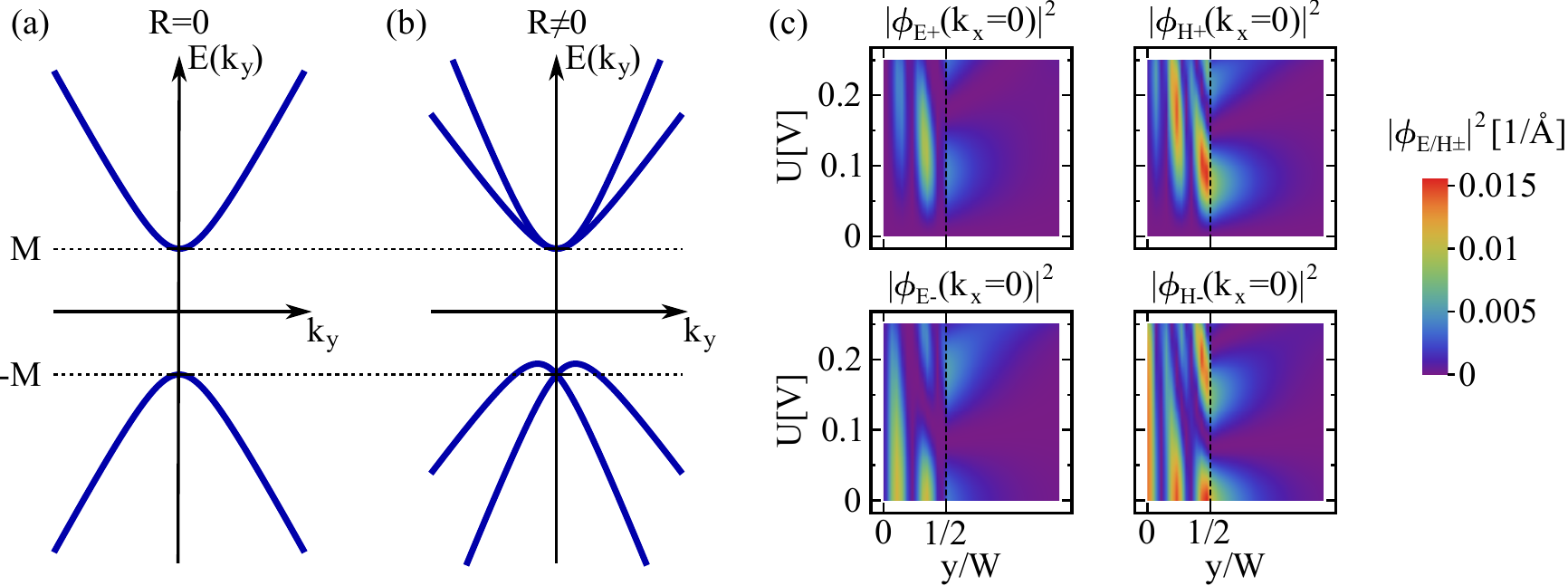}
\caption{Band dispersion at $k_x = 0$ for the 2D quantum well Hamiltonian (a) without Rashba coupling and (b) with Rashba coupling.  The important distinction between the two cases is that there are generically two distinct wavevectors yielding the same energy in (a) but \emph{four} in (b).  Upon gate-defining a wire, confined wavefunctions are then built from additional harmonics when Rashba coupling is present.  The added harmonics result in oscillatory behavior of the confined wavefunctions that, in turn, produce oscillations in physical quantities for the gate-defined wire such as the effective $g$-factors and characteristic Rashba spin-orbit energies. (c) Evolution of the probability amplitudes $|\phi_{E/H\pm}(k_x=0,y)|^2$ for the $n=5$ band as a function of the voltage drop $U$.  Data correspond to a $d = 70$\AA~quantum well hosting a wire of width $W = 800$\AA~and confinement depth $V = 0.06$eV, and were obtained numerically as described in Sec.~\ref{subsec:num}.  The oscillations in the wavefunction tails shown in each plot underlie the oscillatory behavior of physical quantities noted above.}
\label{fig:cone_rashba}
\end{figure*}

The existence of these oscillations can be anticipated based on the following argument.  Without Rashba coupling, the Dirac dispersion for the 2D quantum well along $k_y$ with $k_x = 0$ is sketched in Fig.~\ref{fig:cone_rashba}(a). For a given energy there exists only a single pair of wavevectors $\pm k_y$, so that within the central region confined states are built from plane-waves $e^{\pm i k_y y}$.  Switching on Rashba coupling splits the 2D bands as in Fig.\ \ref{fig:cone_rashba}(b) and changes the situation qualitatively.  In particular, two distinct pairs of wavevectors $\pm k_{y1}$ and $\pm k_{y2}$ now yield the same energy---hence confined wavefunctions involve superpositions of two harmonics, $e^{\pm i k_{y1}y}$ and $e^{\pm i k_{y2}y}$.  The difference in these wavevectors increases with $U$, \emph{i.e.}, $k_{y1}-k_{y2} \propto U$.  Consequently, varying $U$ changes the profile of the confined wavefunctions in an oscillatory fashion.  This effect is visible in Figure \ref{fig:cone_rashba}(c), which displays the numerically computed probability amplitudes $|\phi_{E/H\pm}(k_x = 0,y)|^2$ versus $U$ assuming parameters specified in the caption.  Oscillations in the confined wavefunction tails---which are clearly seen in the figure---in turn produce oscillations in the effective spin-orbit energies (and other physical properties as we will see) characterizing the gate-defined wire.  

The above qualitative argument for the appearance of oscillations relies on the existence of two harmonics {\it inside} the central region where the wire exists. In contrast, the effect of the Rashba coupling outside the wire is of minor relevance for the physical properties of the confined states. This circumstance allows one to simplify the experimental setup of our device with little physical consequence.  In particular, replacing the bottom gates in Fig.\ \ref{fig:setup_hgte}(a) by a single gate as shown in Fig.\ \ref{fig:setup_hgte}(b) leaves independent control over the global chemical potential, the confinement depth $V$, and the voltage drop inside the wire.  One  merely sacrifices independent tunability of the Rashba coupling outside of the wire---which in any case is unimportant.  

Quantitatively capturing such effects clearly requires a more exact treatment of Rashba coupling.  In Sec.\ \ref{subsec:num} we expose the oscillations using exact numerical simulations of the gate-defined wire.  There we show that the effective Rashba energy scale for confined subbands can be tuned from zero to a maximum of a few tens of Kelvin, and back down to zero with a moderate increase in $U$ (see lower panels of Fig.\ \ref{fig:g_eso}).  This level of tunability is highly attractive for spintronics and other applications.  

To close this subsection we remark that the Rashba oscillations discussed above are \emph{not} special to Dirac systems such as HgTe.  Indeed, our qualitative explanation merely required the existence of multiple harmonics, which would arise even in a conventional parabolic dispersion with spin-orbit splitting.  Related non-perturbative phenomena have been explored in conventional semiconductors in Refs.\ \onlinecite{mireles01,governalea04,knobbe05,zhang09}.

\subsubsection{Effective Zeeman splitting}
\label{subsec:g}

Let us now introduce an applied magnetic field and investigate the effective Zeeman splitting imparted to the gate-defined wire.  We focus throughout on magnetic fields ${\bf B} = \mathcal{B}\hat{\bf z}$ directed perpendicular to the quantum well since this orientation yields the strongest effect by far.  The wire's effective Zeeman splitting derives from two physically distinct contributions---the ordinary (spin-orbit-enhanced) Zeeman effect as well as a component due to the orbital part of the magnetic field.  The latter is often justifiably neglected in treatments of wires.  Here, however, we show that the orbital contribution \emph{dominates} as a consequence of the Dirac spectrum exhibited by HgTe.  Initially we treat the case without Rashba coupling, but discuss the strong interplay between Rashba effects and Zeeman splitting at the end of this subsection.  

Consider first the standard Zeeman term for the 2D quantum well,\cite{koenig08}
\begin{equation}
  H_Z = \frac{\mu_{\text{B}}\mathcal{B}}{2}\int d^2{\bf r}\Psi^\dagger
  \left(\begin{array}{cccc}
    g_E                                         &0 & 0 &0\\
    0                                         & g_H &0                                          &0\\
    0 &0 &-g_E                                          &0\\
    0                                         &0 &0                                          &-g_H
  \end{array}\right)\Psi,
  \label{ham_zeeman}
\end{equation}
which contains $g$-factors $g_{E/H}$ for the $E$ and $H$ sectors.  Equation (\ref{projection}) once again allows us to project onto the confined bands of interest.  Taking the weak-field limit where interband mixing is negligible and expanding to leading order in $k_x$, we obtain
\begin{equation}
  H_Z \rightarrow H_{1D,Z} \approx \int_{k_x} \sum_{n} \frac{g_{n,Z}\mu_{\rm B} \mathcal{B}}{2}\psi_n^\dagger\sigma^z \psi_{n},
\end{equation}
with 
\begin{equation}
g_{n,Z} = \int dy  
\Phi_{n+}^T(k_x= 0,y)
\Bigg(\begin{array}{cc}
g_E & 0\\
0      & g_H
\end{array}\Bigg)
\Phi_{n+}(k_x =0,y)\,.\label{int_zeeman}
\end{equation}
Following our usual approach we estimate this component of the effective $g$-factor by considering a quantum well of thickness $d = 70$\AA, for which $g_E=22.7$ and $g_H=-1.21$ according to Table~\ref{table}.  Given these values one generally expects $g_{n,Z}$ to be of order 10---still enhanced compared to the bare electron $g$-factor but much smaller than that for, say, an InSb wire.\cite{nilsson09,mourik12}  Fortunately, as noted earlier the usual Zeeman term constitutes a subdominant contribution to the wire's effective $g$-factor.

Since we are dealing with quasi-1D states possessing strong spin-orbit coupling, the orbital part of the magnetic field masquerades as an \emph{effective} Zeeman splitting for the confined bands.  To incorporate orbital effects we select Landau gauge for the vector potential and replace $\partial_x \rightarrow \partial_x-i e \mathcal{B} y/\hbar$ in the 2D quantum well Hamiltonian in Eq.\ (\ref{general_ham}).  (We continue to set $B = D = 0$ for simplicity, so only the $A$ term is affected by this substitution.)  The orbital contribution to the wire's effective $g$-factor can be similarly obtained by projecting the orbital terms onto the confined bands using Eq.\ (\ref{projection}), neglecting interband couplings as appropriate for weak fields, and expanding to leading order in $k_x$.  This procedure yields the following term in our effective 1D Hamiltonian,
\begin{equation}
  H_{1D,orb} =  \int_{k_x} \sum_{n} \frac{g_{n,orb}\mu_{\rm B} \mathcal{B}}{2}\psi_n^\dagger\sigma^z \psi_{n},
\end{equation}
where
\begin{equation}
  g_{n,orb} = -\frac{4A m_e}{\hbar^2}\int dy y\Phi_{n+}^T(k_x = 0,y)\sigma^x \Phi_{n+}(k_x = 0,y).
  \label{gnorb}
\end{equation}

Appendix~\ref{appendix2} demonstrates that the above integral can be performed exactly, yielding 
\begin{equation}
  \int dy  y
  \Phi_{n,+}^T(0,y)\sigma^x \Phi_{n,+}(0,y)
  =\frac{A}{2M}
\label{analyt_int}
\end{equation}
which, remarkably, is independent of the band index, wire width $W$, and confinement depth $V$.  It is illuminating to express the final result in terms of the \emph{two-dimensional} effective mass for carriers in a uniform HgTe quantum well, $m_{2D}^* = \hbar^2|M|/A^2$.  Upon dropping the irrelevant band index label we obtain
\begin{equation}
  g_{orb} = -2{\rm sgn}(M)\frac{m_e}{m_{2D}^*}.
  \label{gorb}
\end{equation}
Notice that as $M$ goes to zero, orbital effects produce a \emph{divergent} effective $g$-factor; this is reminiscent of the divergent diamagnetic response for gapless Dirac systems such as graphene (see, \emph{e.g.}, Ref.\ \onlinecite{koshino07}).  According to Table \ref{table}, a $d = 70$\AA~thick quantum well is characterized by a very light 2D effective mass $m_{2D}^* \approx 0.0057m_e$---in turn leading to a giant effective $g$-factor $g_{orb} \approx 350$ that greatly exceeds the Zeeman contribution as claimed.    

Several comments are in order.  $(i)$ References \onlinecite{koenig07} and \onlinecite{koenig08} previously emphasized the importance of orbital effects on the $g$-factor for quasi-1D states in HgTe quantum wells, but in the context of quantum spin Hall edge states.  The physics in the two cases is similar but not identical.  In particular, bulk inversion asymmetry terms (which are absent in our treatment) are essential for the effect in the quantum spin Hall case\cite{koenig08}.  $(ii)$ The Dirac dispersion, strong spin-orbit coupling, and small gap provide the key ingredients underlying the giant $g$-factor captured above.  It is the Dirac structure that allows for a finite correction linear in $\mathcal{B}$, which can be seen from Eq.\ (\ref{gnorb}) together with the symmetry properties of the Dirac wavefunctions in Eq.\ (\ref{PhiProperties}).  Spin-orbit coupling intrinsic to the 2D quantum well Hamiltonian ensures that this linear correction lifts Kramer's degeneracy, and the small gap guarantees that this happens very efficiently.  $(iii)$  It is instructive to contrast our results for HgTe with the behavior for gate-defined wires in systems exhibiting conventional parabolic dispersion, \emph{e.g.}, an electron-doped GaAs quantum well.  There the analogue of Eq.\ (\ref{gnorb}) would vanish by symmetry if the confinement potential is symmetric under $y\rightarrow-y$, so that the leading perturbative orbital effect appears at second order in $\mathcal{B}$.  A linear term at $k_x \neq 0$ could still arise if the wire forms from asymmetric confinement, but such a term would \emph{not} manifest as an effective $g$-factor for the confined bands since spin degeneracy would remain unbroken (at least in the absence of spin-orbit coupling).  

So far in our discussion of effective Zeeman splitting we have entirely neglected Rashba spin-orbit interactions induced by a voltage drop $U$ across the quantum well width.  The results above still hold in the perturbative limit where Rashba coupling is weak, modulo small corrections coming from orbital effects induced by the Rashba terms.  In the non-perturbative regime, however, the interplay between Rashba and orbital magnetic field effects produces still more striking physics.  As described in Sec.\ \ref{subsec:eso} large voltage drops generate order-one modifications of the confined wavefunctions that, crucially, are oscillatory in $U$.  The effective $g$-factor for the wire arises from projecting the orbital magnetic field terms using these modified wavefunctions, and hence inherits their oscillations.  This effect is by no means small as we demonstrate numerically in the following section and analytically in a simplified model in Appendix \ref{appendix3}.  In fact, as we will see below the effective $g$-factor for the confined subbands is exquisitely tunable and can be adjusted by factors of \emph{several hundred} (and modulated in sign) by varying $U$ over moderate voltage ranges.

\subsection{Numerical results}
\label{subsec:num}

\begin{figure}[t]
\centering
\includegraphics[scale=0.35]{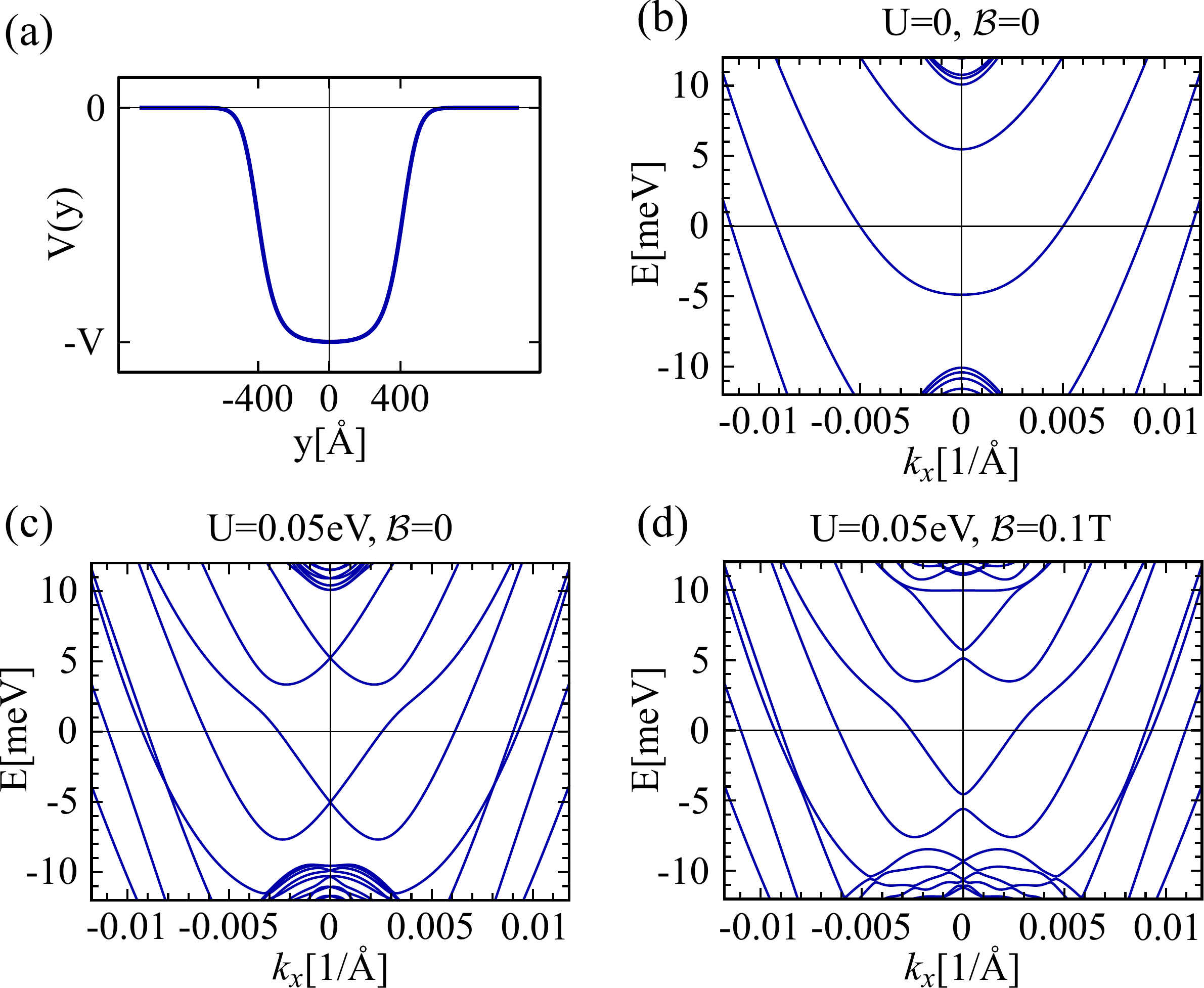}
\caption{(a) Smeared confinement potential used to define a wire in our numerical simulations.  (b)-(d) Calculated band dispersions in various cases assuming a $d=70$\AA~ thick quantum well, wire width $W=800$\AA, and confinement potential depth $V=0.052$eV.  In (b) $U = \mathcal{B} = 0$ so that both Rashba coupling and a magnetic field are absent.  A voltage drop $U = 0.05$V is present in (c) leading to Rashba splitting of the confined bands.  A magnetic field separates the Rashba-split bands, as (d) illustrates for a perpendicular field of strength $\mathcal{B} = 0.1$T.  Dense bands at the top and bottom of these plots represent extended bulk states. }
\label{fig:bands_example}
\end{figure}

Next we complement our analytic treatment above with more accurate numerical simulations of the gate-defined HgTe wire.  The purpose of these numerics is twofold.  First, in the preceding subsections several simplifying assumptions were made to facilitate analytical progress---\emph{e.g.}, focusing on the long-wavelength limit, considering a step-like confinement potential, \emph{etc}.  Here we simulate the full 2D quantum well Hamiltonian $H = H_{2D} + H_{\rm conf} + H_R + H_Z$ [the respective terms are defined in Eqs.\ (\ref{general_ham}), (\ref{Hconf}), (\ref{ham_rashba}), and (\ref{ham_zeeman})] with these assumptions relaxed.  We continue to focus on a $d = 70$\AA~thick quantum well hosting a gate-defined wire of width $W = 800$\AA, but now for the first time include the $B$ and $D$ terms in $H_{2D}$ using the parameter values quoted in Table \ref{table}.  A more experimentally realistic confinement potential $V(y)$, in which the confinement walls are broadened over a distance of $\sim 200$\AA, will also now be taken; for a sketch see Fig.\ \ref{fig:bands_example}(a).  And finally, we properly account for orbital effects of the perpendicular magnetic field ${\bf B} = \mathcal{B}{\bf \hat{z}}$ by sending $\partial_x \rightarrow \partial_x-i e \mathcal{B} y/\hbar$ in all terms in $H$ (including the $B,D$ terms and the Rashba Hamiltonian).  The second and more important goal of our numerics is to quantitatively capture the non-perturbative effects of Rashba coupling alluded to earlier---in particular the prodigious oscillations in the effective Rashba energy and $g$-factor characterizing subbands in the gate-defined wire.  

It is convenient to treat the quantum well as continuous along $x$ (to take advantage of translation symmetry) but discretized along $y$ so that one can describe the system in terms of an effective tight-binding lattice model that is readily simulated.  The discretization is achieved in the standard way.  One first  expresses terms involving $\partial_y$ in momentum space and then replaces $k_y \rightarrow \sin(k_y a)/a$ in the continuum Hamiltonian, where $a$ is the discretized model's lattice spacing.  Finally, a partial Fourier transform $\Psi(k_x,k_y)=\frac{1}{\sqrt{N}}\sum_j e^{-ik_y y_j}\Psi(k_x,y_j)$ results in an effective hopping problem on an $N$-site chain with sites labeled by $y_j$.  In all simulations we take $N = 500$ and assume periodic boundary conditions along $y$ to eliminate unwanted edge effects.\footnote{Far away from the quantum wire (in the $y$ direction) our system contains a narrow region with opposite field direction such that the overall flux vanishes as required in a cylinder geometry. Such a field configuration has a negligible effect, however, on the confined states.}  The lattice spacing is adjusted such that the width of the system along $y$ is $a N = 10 W$; the exponential tails of the confined wavefunctions are then well-resolved numerically.  

Figures \ref{fig:bands_example}(b)-(d) present typical band structures calculated within the above scheme, assuming a confinement depth $V = 0.052$eV.  Case (b) corresponds to $U = \mathcal{B} = 0$ where neither Rashba coupling nor a magnetic field are present.  Hence all bands there are doubly degenerate.  Notice that the $n = 1$ and $2$ subbands overlap with extended bulk states at small $k_x$.  In (c) the voltage drop across the quantum well is set to $U = 0.05$V, resulting in a pronounced Rashba splitting of the $n = 3$ and $4$ subbands.  A perpendicular magnetic field of strength $\mathcal{B} = 0.1$T is present as well in (d) and produces a clear separation between these Rashba-split bands.  
  
In such simulations we quantify the Rashba spin-orbit energy for subband $n$ by considering the $\mathcal{B} = 0$ limit and defining $E_{SO,n} = \frac{1}{4}\hbar \alpha_n k_{F,n}$.  Here the Rashba parameter $\alpha_n$ is deduced from the slope of the dispersion at $k_x = 0$ while $k_{F,n}$ is the Fermi wavevector for subband $n$ when the chemical potential resides at the $k_x = 0$ crossing for that subband [\emph{e.g.}, $\mu \approx -5$meV for $n = 3$ and $\mu \approx 5$ meV for $n = 4$ in Fig.\ \ref{fig:bands_example}(c)].  This definition of $E_{SO,n}$ reproduces our previous expression $\frac{1}{2}m_n \alpha_n^2$ in the case of a simple quadratic dispersion, but is more appropriate when significant non-parabolicities arise as is often the case here.  To extract the magnitude of the effective $g$-factor for subband $n$, denoted $g_n$, we equate the magnetic-field-induced splitting of the confined bands at $k_x = 0$ with $|g_n| \mu_B \mathcal{B}$.  (One can infer the sign of $g_n$ analytically as discussed below.)  For concreteness we use a field strength $\mathcal{B} = 0.005$T for this extraction throughout.  Note that it is difficult to meaningfully compute $E_{SO,n}$ and $g_n$ for bands that intersect extended states at small $k_x$ (\emph{e.g.}, $n = 1,2$ subbands in Fig.\ \ref{fig:bands_example}), so below we will not quote spin-orbit energies and effective $g$-factors for such bands.  

\begin{figure*}[t]
\centering
\includegraphics[scale=0.42]{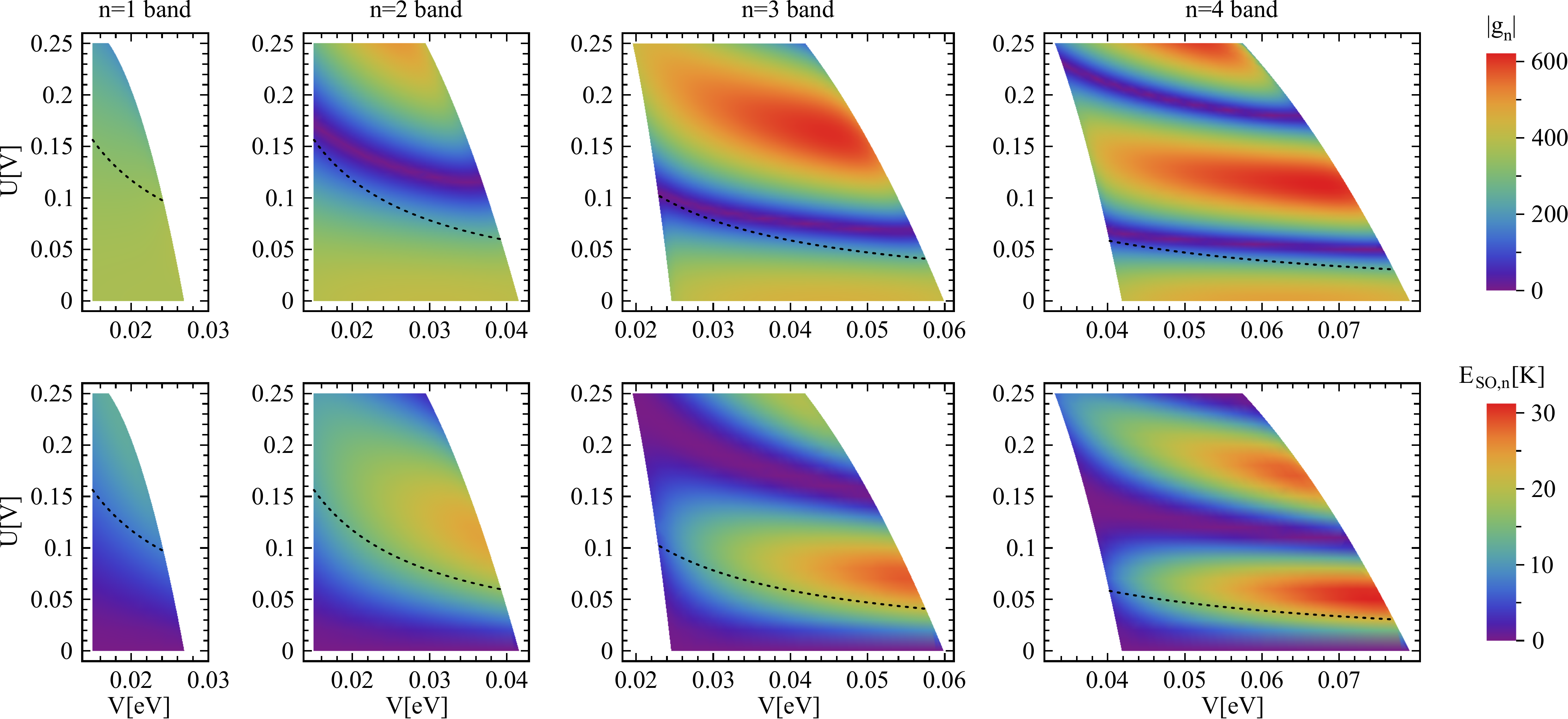}
\caption{Color-scale plot illustrating the remarkable oscillations in the effective $g$-factor (upper row) and spin-orbit energy (lower row) for the four lowest confined subbands.  The horizontal axis represents the depth $V$ of the confinement potential used to define the wire, while the vertical axis corresponds to the potential drop $U$ across the quantum well due to a perpendicular electric field.  Data were extracted from numerical simulations of a $d=70$\AA~quantum well supporting a wire of width $W=800$\AA, assuming a smooth confinement potential.  For a given band, the $g$-factor and spin-orbit energy are well-defined only if that band supports confined states at $k_x=0$. This condition cuts off the plots at small and large $V$. Dashed lines indicate the crossover between perturbative and non-perturbative Rashba coupling regimes according to Eq.~(\ref{perturbative_regime}).}
\label{fig:g_eso}
\end{figure*}

Figure \ref{fig:g_eso} illustrates the dependence of the effective $g$-factor (upper row) and spin-orbit energy (lower row) on the voltage drop $U$ and confinement depth $V$ for the first four confined subbands.  White regions correspond to $U,V$ values where a given confined subband either does not exist, or intersects bulk states at $k_x = 0$ so that the quantities of interest can not be determined as remarked above.  The dashed line in the figures roughly indicates the values of $U$, according to Eq.\ (\ref{perturbative_regime}), where the crossover between perturbative and non-perturbative Rashba coupling regimes transpires.   Consider first the perturbative limit.  At $U = 0$ the $g$-factor ranges from $\sim 370$ for the $n = 1$ band to $\sim 470$ for the $n = 4$ band.  Such enormous values again arise because of orbital magnetic field effects, with deviations from our previous analytical estimate arising primarily from the $B$ and $D$ terms.  The attainable Rashba energies in this regime are of order 10K in agreement with estimates from Sec.\ \ref{subsec:eso}.  

Far more interesting is the non-perturbative limit where oscillations in both quantities are visible.  The following key features are worth highlighting.  The oscillation frequency increases with the band index, which is why the variation with $U$ over the range shown is relatively minor for the $n = 1$ band but becomes increasingly pronounced in the higher subbands.  More importantly, moderate changes in $U$ effect \emph{giant} modulations in both the effective $g$-factor magnitudes and Rashba energies---the former varying from zero to more than 600, the latter between 0 and $\sim 30$K.\footnote{Note that having $E_{SO,n} = 0$ does \emph{not} mean that effects of Rashba coupling on the energy spectrum for confined states is absent.  Rather, this condition implies that the leading Rashba term is cubic in $k_x$ rather than linear.}  Both quantities oscillate with roughly the same period but do so out of phase.  That is, for a given subband the $g$-factor amplitude reaches a maximum when the spin-orbit energy is minimized and vice versa.  

Appendix \ref{appendix3} treats a simplified model that allows one to analytically capture the main features of these oscillations for high subbands.  As the calculation is somewhat lengthy we will not comment on the details here but instead simply note two important conclusions.  First, our analytical study reveals that for high subbands the oscillation period in $U$ is approximately given by
\begin{equation}
  \Delta U \approx \frac{4\pi A^2}{eFVW}.
  \label{DeltaU}
\end{equation}
Notice that the perturbative regime identified in Eq.\ (\ref{perturbative_regime}) persists to roughly one quarter of a wavelength of the oscillations.  And second, Appendix \ref{appendix3} demonstrates that the lines in Fig.\ \ref{fig:g_eso} at which $|g_n|$ vanishes are associated with sign changes for $g_n$.  Thus the effective $g$-factors for the confined subbands are highly tunable both in magnitude \emph{and} sign.  

It is important to address how the oscillations depend on the width $W$ of the wire.  On one hand Eq.\ (\ref{DeltaU}) illustrates that the oscillation period decreases with $W$.  But on the other, a shortened period cuts off the quadratic rise of the spin-orbit energy with $U$ in the perturbative regime [see Eq.\ (\ref{E_SO_perturbative})] at a reduced value of $U$.  The net effect is that wider wires yield smaller attainable Rashba energies.  Consequently, if one desires to maximize the effective spin-orbit coupling, narrow wires are generally advantageous.  We have confirmed numerically, however, that the magnitude of the $g$-factor oscillations remains roughly constant upon increasing $W$---at least up to values $W \approx1500$\AA.  This is perhaps not too surprising since in the perturbative regime Eq.\ (\ref{gorb}) shows that the dominant orbital contribution to the $g$-factor is largely insensitive to both $U$ and $W$ (unlike Rashba coupling).  Measuring the giant $g$-factor oscillations experimentally, with magnitudes peaking at $\sim 600$, should thus be even easier in wider wires.  

In summary, gate-defined HgTe wires possess the fascinating property that their subband-dependent $g$-factors and spin-orbit energies can both be tuned continuously over enormous ranges simply by changing the gate voltage. However, since these parameters vary out of phase, one cannot maximize both simultaneously. Values of $U$ leading to a `compromise' where both quantities remain large are still possible, however, and we shall exploit such cases in the next section when discussing one particularly appealing potential application---the pursuit of Majorana fermions.

\section{Majorana zero-modes in gate-defined ${\rm HgTe}$ wires}
\label{sec:pairing}

When a wire with an odd number of channels acquires a bulk Cooper-pairing gap, the system can form a topological superconducting state supporting protected Majorana zero-modes at its endpoints.  The physics is intimately related to that of the Kitaev chain introduced in Ref.\ \onlinecite{kitaev01}.  A particularly powerful means of fashioning such a setup experimentally was proposed by Lutchyn \emph{et al}.\cite{lutchyn10} and Oreg \emph{et al}.\cite{oreg10} (see Refs.~\onlinecite{potter10,lutchyn11,stanescu11,wimmer10,potter11_2} for multichannel generalizations).  These authors showed that a topological phase can be engineered in spin-orbit-coupled wires that are subjected to a magnetic field and proximity coupled to an ordinary $s$-wave superconductor.  The magnetic field opens up chemical potential windows where an odd number of channels are occupied as desired.  Spin-orbit coupling meanwhile causes the spin to depend nontrivially on momentum in each partially occupied band---allowing an $s$-wave order parameter to open a full pairing gap even in such odd-channel regimes.  

Ideally, wires featuring both large $g$-factors and spin-orbit energies are desirable for this proposal.  The former permits one to operate at relatively weak magnetic fields---hence disturbing the parent superconductor weakly---while the latter (among other benefits) leads to enhanced robustness against disorder as discussed below.  It is interesting to explore the formation of Majoranas in gate-defined HgTe wires since they offer the possibility of satisfying both criteria simultaneously.  A superconducting proximity effect can be induced in the HgTe wire using a setup similar to Fig.\ \ref{fig:setup_hgte}(c) in which the central top gate is replaced by an $s$-wave superconductor.  Such a configuration does not allow independent tuning of the the electron density and Rashba coupling for the wire, though it is conceivable that one can enhance the tunability by, say, employing additional top gates adjacent to the superconductor.  In any case we assume in our analysis below that Rashba coupling strengths and densities similar to those captured in the previous section can be realized here as well.  We will first treat the clean case by studying numerically the full 2D quantum well Hamiltonian with proximity-induced Cooper pairing, and then discuss disorder effects within a simplified effective 1D Hamiltonian.

\subsection{Numerical phase diagram}
\label{subsec:phase_diag}

It is useful to first explore the rough phase diagram, and achievable gaps in the topological regimes, for the device in Fig.\ \ref{fig:setup_hgte}(c) assuming the clean limit.  To this end we follow the methods outlined in Sec.\ \ref{subsec:num} to exactly diagonalize the 2D quantum well Hamiltonian $H = H_{2D} + H_{\rm conf} + H_R + H_Z + H_\Delta$ in a perpendicular magnetic field ${\bf B} = \mathcal{B}{\bf \hat{z}}$.  The first four terms were previously simulated in Sec.\ \ref{subsec:num} and reflect the kinetic energy for the quantum well, smooth confinement potential defining the wire, Rashba coupling, and Zeeman splitting---including orbital magnetic field contributions where appropriate.  The last term $H_\Delta$ encodes (crudely) the Cooper pairing inherited from the neighboring superconductor.  We model this term by pairing opposite pseudospins from the $E$ and $H$ sectors with pairing strengths $\Delta_E$ and $\Delta_H$,
\begin{equation}
  H_\Delta=\int d{\bf r}(\Delta_E\psi_{E+}\psi_{E-}+\Delta_H\psi_{H+}\psi_{H-}+\text{H.c.})\,.\label{ham_pairing}
\end{equation}
For simplicity, we have assumed spatially uniform $\Delta_{E/H}$ above and neglected other symmetry-allowed pairing terms.  Although we will simulate the full 2D Hamiltonian, it is useful to note that one can ascertain the effect of proximity-induced pairing on the wire by projecting Eq.\ (\ref{ham_pairing}) onto the confined subbands.  Such a projection produces both intra- and inter-band pairing terms whose magnitudes depend on the wavefunctions and the precise values of $\Delta_{E/H}$.\footnote{In contrast to the effective Rashba coupling and $g$-factors for the wire, the effective pairing amplitudes do \emph{not} oscillate with $U$.  The distinction arises because the $y$-integration of the projected pairing terms is dominated by $|y|\lesssim W/2$; hence they are weakly sensitive to the oscillatory confined wavefunction tails.}  At zero magnetic field time-reversal symmetry dictates that both $\Delta_E$ and $\Delta_H$ are real but fixes neither their relative amplitudes nor signs.  Absent detailed microscopic modeling which we will not attempt here, we set 
\begin{equation}
  \Delta_E = \Delta_H \equiv \Delta
\end{equation}
throughout to minimize the number of free parameters.\footnote{Note that if $\Delta_E$ and $\Delta_H$ have similar magnitudes but opposite signs, a very small proximity-induced gap may arise due to cancellations.  Such a situation is non-generic, however, and is not expected to be relevant in practice.}  All numerics discussed below were carried out for an infinite $W = 800$\AA~wire using parameters from Table \ref{table}.

\begin{figure*}[t]
\centering
\includegraphics[scale=0.8]{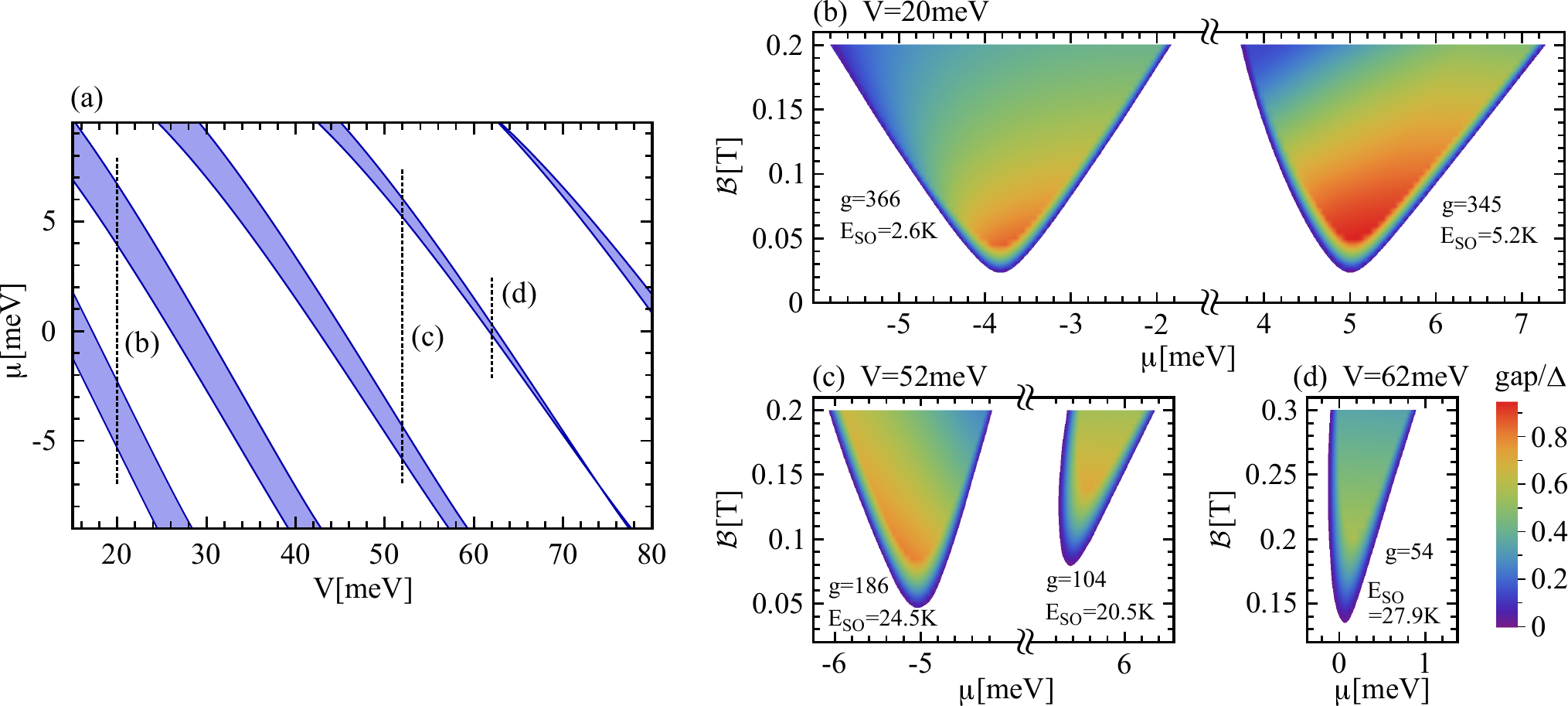}
\caption{(a) Windows of chemical potential $\mu$ (blue regions) in which the gate-defined wire exhibits an odd number of channels, assuming a voltage drop $U = 0.05$V and perpendicular magnetic field $\mathcal{B} = 0.15$T.  The horizontal axis denotes the confinement depth $V$ defining the wire.  Inside of these windows, introducing proximity-induced superconductivity yields a topological phase supporting Majorana zero-modes at the wire's ends.  (b)-(d) Phase diagrams in the $\mathcal{B}-\mu$ plane along the constant-$V$ cuts labeled by dashed lines in (a).  Shaded areas denote topological phases, with the bulk gap (in units of the pairing energy $\Delta = 0.25$meV) indicated by the color scale.  For each topological region we also display the topmost partially occupied subband's effective $g$-factor and spin-orbit energy---both of which are quite large in all cases.}
\label{fig:phase_diag}
\end{figure*}

Figure \ref{fig:phase_diag}(a) illustrates the chemical potential windows (blue regions) as a function of the confinement depth $V$ where the wire possesses an odd number of channels as required for topological superconductivity.  The data correspond to $\Delta = 0$, $\mathcal{B} = 0.15$T, and a voltage drop $U = 0.05$V.  The chemical potential resides within the magnetic-field-induced gap in the $n = 1$ band at the leftmost strip, the $n = 2$ band at the next strip over, \emph{etc.}  Oscillations in the band-dependent effective $g$-factors account for the varying width of these strips; recall Fig.\ \ref{fig:g_eso}.  

Upon turning on $\Delta$, a topological phase supporting Majorana modes appears beyond a critical magnetic field in these odd-channel regimes.  Figures \ref{fig:phase_diag}(b) through (d) show $\mathcal{B}-\mu$ phase diagrams at constant $V$ cuts [dashed lines in Fig.\ \ref{fig:phase_diag}(a)] using $\Delta = 0.25$meV.  The shaded regions represent topological phases, the boundaries of which correspond to the fields that close the bulk gap.  As in other wire setups\cite{lutchyn10,oreg10}, the minimum required field follows from $E_{\rm Zeeman} \approx \Delta$, where $E_{\rm Zeeman} = \frac{1}{2}|g_n| \mu_B \mathcal{B}$ is the Zeeman energy for the topmost partially occupied band.  (Roughly, this is the field required to overcome interband pairing.)  Beneath each topological region we also list the effective $g$-factor and spin-orbit energy characterizing the uppermost band.  These quantities are encouragingly large in all plots---for reference one may compare to electron-doped InSb wires for which $g \approx 50$ and $E_{SO} \approx 1$K.  The small field scale at which the topological phase sets in [$\sim 25$mT in (b)] is also noteworthy considering the sizable $0.25$meV pairing energy assumed.  

The magnitude of the bulk gap in the topological phase is indicated by the color scale in Figs.\ \ref{fig:phase_diag}(b)-(d).  Near the phase boundaries with the trivial state, $k_x = 0$ excitations always determine the minimal gap.  In the interior of the topological regimes, however, the gap is set by finite-$k_x$ excitations near one of the Fermi points.  Somewhat counterintuitively, the minimum excitation energy here need not be set by the topmost partially occupied band.  Rather, in some cases `background' confined subbands yield the minimum-energy gap.  This indeed occurs in Figs.\ \ref{fig:phase_diag}(c) and (d) and is responsible for the generally smaller gaps present there in comparison to Fig.\ \ref{fig:phase_diag}(b).  To illustrate the physics, we note that cut (c) corresponds to the band structure displayed in Fig.~\ref{fig:bands_example}(d) where two `background' subbands cross at $E \approx -4$meV.  Interband pairing becomes appreciable near that crossing and conspires to reduce the gap somewhat.  Such effects are likely non-generic but should be kept in mind.  

\begin{figure}[t]
\centering
\includegraphics[scale=0.88]{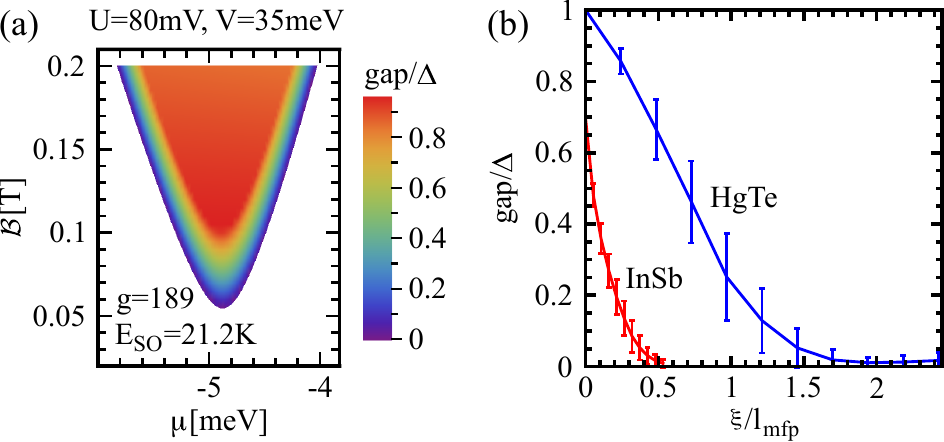}
\caption{(a) Phase diagram for a HgTe wire with voltage drop $U=80$mV, confinement depth $V=35$meV, and pairing strength $\Delta=0.25$meV.  The color scale indicates the gap for the topological phase.  The combination of a large gap, $g$-factor, and spin-orbit energy render this regime highly favorable experimentally.   In particular, the large spin-orbit energy implies reduced sensitivity to disorder.  (b) Gap as a function of the ratio of the superconducting coherence length and the mean free path $\xi/l_{\rm mfp}$, obtained by solving an effective disordered single-band 1D model for the topological phase realized in (a).  The data correspond to a magnetic field $\mathcal{B} = 0.2$T and effective pairing amplitude $\Delta_{\rm eff}=0.1$meV, with the chemical potential situated in the center of the Zeeman gap.  The upper curve corresponds to HgTe parameters relevant for (a), while the lower curve corresponds to InSb wire parameters.  The larger gap for the former stems from the greatly enhanced spin-orbit energy.}
\label{fig:disorder}
\end{figure}

A still more favorable experimental situation appears in Fig.\ \ref{fig:disorder}(a).  The data here correspond to a larger voltage drop of $U = 0.08$V and a confinement depth $V = 0.035$eV, with the $n = 1$ and 2 subbands partially occupied.  While the large effective $g$-factor and spin-orbit energy for the uppermost confined band are comparable to those in Fig.\ \ref{fig:phase_diag}(c), the gap protecting the topological phase is significantly larger and decays much more slowly with the magnetic field over the interval shown.  The enhanced robustness follows simply because `background' confined subbands do not limit the gap here [in contrast to Fig.\ \ref{fig:phase_diag}(c)].

\subsection{Disorder effects}

Lastly we discuss crucial effects of disorder on the topological phase.  In the presence of time-reversal symmetry, Anderson's theorem dictates that random potential disorder does not degrade the superconducting gap for $s$-wave-paired systems.\cite{anderson59,potter11}  This theorem does not apply, however, to the topological phase since its formation requires a finite magnetic field.  The severity of the gap's degradation by disorder depends on the degree to which time-reversal symmetry has been broken.  A useful way to quantify this is via the ratio of the Zeeman and spin-orbit energies, $E_{\rm Zeeman}/E_{SO}$, for the highest partially occupied confined band.  Working in the limit $E_{\rm Zeeman}/E_{SO} \ll 1$ is highly advantageous since here spins at the Fermi level feel the effects of the field only marginally.  In this sense time-reversal symmetry is weakly violated, imparting the system with greater immunity against disorder.\cite{potter11,sau12}

At the minimum fields required to access the topological states shown in Figs.\ \ref{fig:phase_diag} and \ref{fig:disorder}, the Zeeman energy falls in the range $E_{\rm Zeeman} \sim 2-3$K (roughly the size of $\Delta$).  This rough scale together with the spin-orbit energies listed in the figures suggest that it is indeed possible to stabilize Majoranas in the coveted spin-orbit-dominated regime $E_{\rm Zeeman}/E_{SO} \ll 1$.  (Since the topological phase requires $E_{\rm Zeeman} \gtrsim \Delta$, one can always trivially access this regime by making the proximity effect poor.  The key point is that for HgTe wires this remains feasible even with generous values of $\Delta$ and gaps exceeding 1K.)  The parameters for Fig.\ \ref{fig:disorder}(a) appear particularly promising due to the large attainable gap.  

Next we provide a rough illustration of the advantage afforded by the large spin-orbit energies found above by modeling a gate-defined HgTe wire by an effective single-band 1D Hamiltonian with random potential disorder.  We caution that our results here are only meant to expose general trends.  For one, disorder in the 2D quantum well will generate randomness in quantities aside from the local potential.  The neglect of other subbands is also certainly crude since the most promising cases identified above correspond to multi-channel situations.  To mitigate the effects of `background' subbands we will use parameters relevant for the topological phase in Fig.\ \ref{fig:disorder}(a) since there additional bands are at least unimportant in the clean limit.  

With these caveats in mind, consider the 1D Hamiltonian $H = H_{0} + H_{\rm disorder}$, where
\begin{eqnarray}
  H_0 &=& \int dx \psi^\dagger\left(-\frac{\hbar^2\partial_x^2}{2m}-\mu - i \hbar\alpha \sigma^y \partial_x + \frac{1}{2}g\mu_B \mathcal{B}\sigma^z\right)\psi
  \nonumber \\
  &+& \int dx \Delta_{\rm eff}(\psi_+ \psi_- + H.c.)
\end{eqnarray}
describes the clean wire with proximity-induced pairing and $H_{\rm disorder}$ encodes the random potential.  We choose parameters for $H_0$ to reproduce quantities relevant for Fig.\ \ref{fig:disorder}(a) at $\mathcal{B} = 0.2$T ($\frac{1}{2} m\alpha^2 = 21.2$K, $\hbar\alpha = 2.2$eV\AA, and $g = 165$, which is slightly reduced from the value quoted in Fig.\ \ref{fig:disorder}(a) due to nonlinear effects).  The Hamiltonian is most easily simulated upon mapping the problem onto a discretized lattice model (here we typically use 8000 lattice sites).  One can then implement the random potential as
\begin{equation}
H_{\text{disorder}}=\sum_{x}V_x\psi^{\dagger}_x\psi_x\,
\end{equation}
with $x$ now labeling discrete lattice sites.  We choose the disorder potential $V_x$ to exhibit Gaussian white noise correlations with $\overline V_x = 0$ and $\overline{V_xV_{x'}}=\delta_{x,x'}\mathcal{W}^2$.  The disorder strength $\mathcal{W}$ can be related to the mean-free path $l_{\rm mfp} =v_{\text{F}}\tau$ via $\tau^{-1}=\frac{2\pi}{\hbar}\mathcal{W}^2aN(\epsilon_{\text{F}})$, where $a = 6.46$\AA~which is the HgTe lattice constant and $v_{\text{F}}$ and $N(\epsilon_{\text{F}})$ respectively denote the clean-system Fermi velocity and density of states at the Fermi energy. 

We have performed simulations of $H$ with various disorder realizations in the case $\Delta_{\rm eff} = 0.1$meV, $\mathcal{B}=0.2$T, and $\mu = 0$, corresponding to the topological region of Fig.~\ref{fig:disorder}(a) with the chemical potential lying at the center of the Zeeman gap for the topmost band.\footnote{A smaller value of $\Delta_{\rm eff}$ is chosen here simply so that with InSb parameters the topological phase does not appear too close to the critical point at a field of $\mathcal{B} = 0.2$T.}  The upper curve of Fig.\ \ref{fig:disorder}(b) displays the resulting disorder-averaged gap as a function of the ratio of the superconducting coherence length and the mean free path $\xi/l_{\rm mfp}$, where $\xi=\pi v_{\text{F}}\hbar/\Delta_{\text{eff}}$.\cite{potter11}  Also shown for comparison are results obtained with InSb wire parameters---$m = 0.015m_e$, $\hbar \alpha = 0.23$eV\AA, $g = 50$---using the same magnetic field, chemical potential, and pairing amplitude (note that the InSb lattice constant $a=6.48$\AA~is almost identical to that of HgTe).  Due to the smaller ratio $E_{\rm Zeeman}/E_{SO}$, for HgTe the gap takes on nearly the full value of $\Delta_{\rm eff}$ in the clean case and remains substantially larger than that of InSb at comparable $\xi/l_{\rm mfp}$.\footnote{It is tempting to use measured values for mean free paths for HgTe quantum wells and InSb wires to attempt a further comparison.  We avoid this since the relevant quantities in the presence of a superconductor are not known.} This is consistent with the qualitative argument discussed above.  Of course what matters ultimately is whether one can successfully induce a proximity effect and access the topological phase experimentally in the first place.  The results above (which again should only be taken as a rough guide) further suggest that pursuing Majorana fermions in gate-defined HgTe wires is a worthwhile endeavor.

\section{Discussion}\label{sec:conclusion}

Gate-defined HgTe wires, for a number of reasons, offer great potential for the pursuit of applications requiring the manipulation of electronic spin degrees of freedom.  The parent quantum wells can be fabricated with quite high mobility (at least up to $1.5\times10^5$cm$^2$V$^{-1}$s$^{-1}$)\cite{koenig07}.  Confined subbands in the gate-defined wire can exhibit giant effective $g$-factors (exceeding 600!) and large Rashba spin-orbit energies measuring tens of Kelvin.  Even more striking is the exceptional tunability of these quantities evident in Fig.\ \ref{fig:g_eso}---both can be altered from the large values quoted above through zero in an oscillatory fashion by moderate variations in gate voltages.  

It is useful to summarize the origin of these effects.  The small gap and large intrinsic spin-orbit coupling for HgTe together cause the \emph{orbital} part of the magnetic field to enormously enhance the effective $g$-factors for the wire.  Such orbital contributions would typically be negligible in weakly spin-orbit coupled systems but dominate the Zeeman splitting here.  As for the oscillatory behavior in the wire's effective Rashba coupling and $g$-factors, these are rooted in non-perturbative modifications of the confined wavefunctions by a gate-induced perpendicular electric field.  In principle wires patterned in more conventional electron-doped quantum wells (\emph{e.g.}, GaAs) can also undergo such oscillations, at least in the Rashba coupling, though likely with much smaller amplitude.  Similar physics is in fact implicitly present in the wide wires studied in Refs.\ \onlinecite{mireles01,governalea04,knobbe05,zhang09}.  Armed with these insights, we suggest that gate-defined wires in hole-doped quantum wells may exhibit very similar physics to those in HgTe.  This would be interesting to explore in greater detail in future work.  

As one enticing application, we explored the prospect of employing HgTe wires to stabilize Majorana zero-modes.  The giant accessible $g$-factors and large spin-orbit energies lead to a number of potential advantages, notably the ability to access a topological superconducting state at quite small fields (as low as tens of milliTeslas) and with a sizable gap that exhibits reduced sensitivity to disorder.  Here we wish to comment further on additional advantages offered by gate-defined wires, which apply not just to HgTe-based structures but to any suitable two-dimensional electron gas.  In particular, the formation of wire networks appears to be relatively straightforward in this class of systems, requiring only additional patterning of gates on the quantum well.  

\begin{figure}[t]
\centering
\includegraphics[scale=0.6]{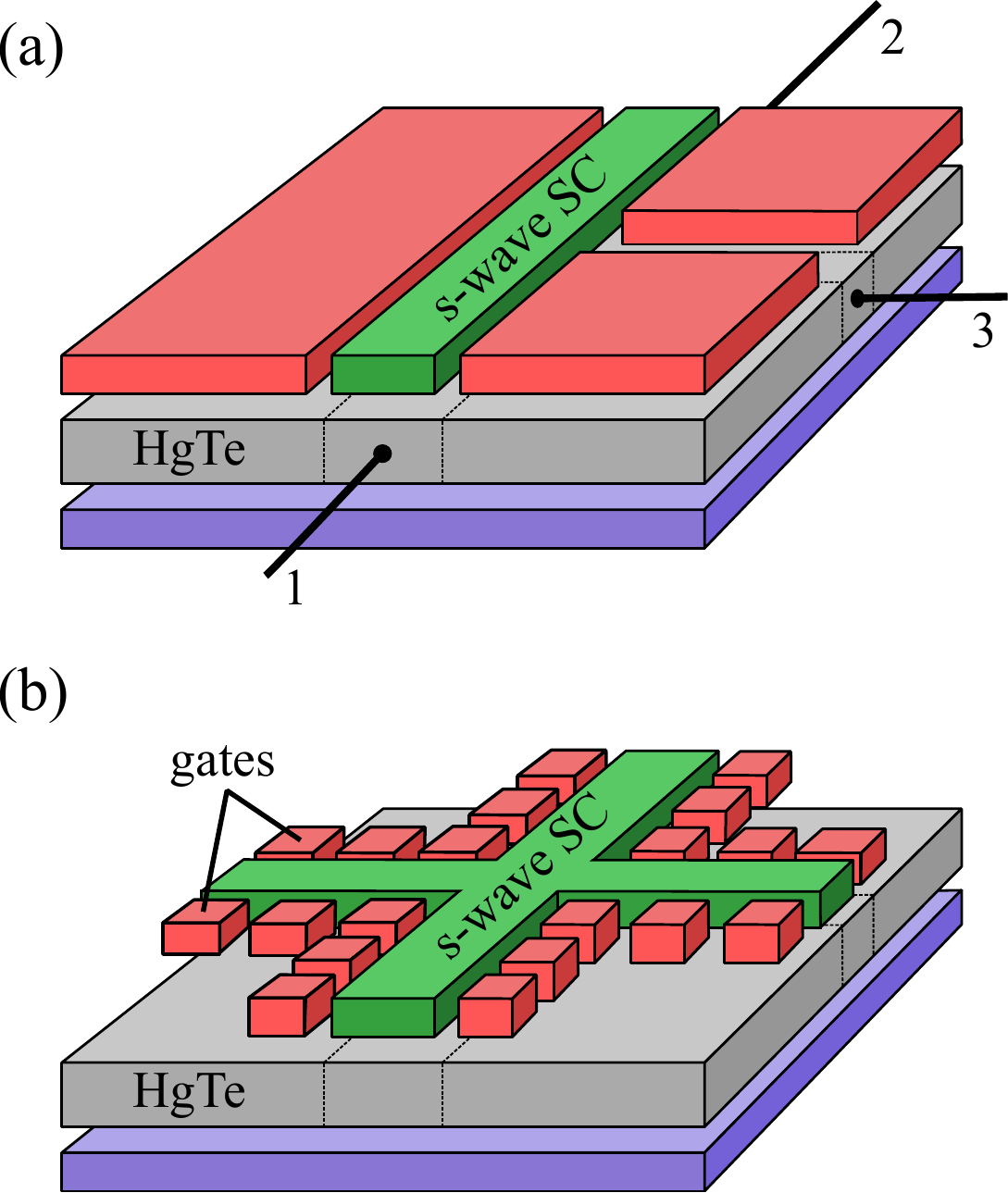}
\caption{(a) Multiterminal transport setup allowing for a refined detection of Majorana zero-modes. A current through leads 1 or 2 allows one to probe the hallmark zero-bias anomaly stemming from Majoranas.  To readily resolve the gap closure (and revival) that accompanies the onset of the zero-bias peak, transport can also be measured from lead 3 which probes the wire's bulk. (b) Braiding of Majorana modes may be performed using a network of superconducting wires that can be fashioned simply by deposition of additional gates on the quantum well.}
\label{fig:exp_outlook}
\end{figure}

There are at least two interesting applications one can envision with such wire networks.  The first is an improved detection scheme for the onset of the topological phase and accompanying Majorana zero-modes via transport.  Consider, for instance, the multi-terminal setup shown in Fig.~\ref{fig:exp_outlook}(a).  Leads 1 and 2 in the figure allow one to inject current into the ends of the superconducting part of the wire to search for the hallmark Majorana-mediated quantized zero-bias anomaly,\cite{sengupta01,bolech07,nilsson08,law09,qu11,fidkowski12} as has been done in recent experiments.\cite{mourik12,das12}  Several authors have pointed out, however, that the closing of the bulk gap at the topological phase transition may be difficult to resolve in such a measurement, because the wavefunctions for the gapless excitations may have very little weight near the ends of the superconductor (hence producing a weak transport signal).\cite{stanescu12,pientka12,prada12}  Measuring transport from lead 3, which impinges on the \emph{bulk} of the superconducting wire segment, should avoid this complication entirely and provide important complementary information about bulk physics.  In particular, observing a collapse and revival of the bulk gap coincident with the appearance of a stable zero-bias peak (even if not quantized) would provide extremely strong evidence for Majorana zero-modes.  Such an experiment should also be able to distinguish `accidental' zero-bias peaks driven by disorder\cite{liu12,bagrets12,pikulin12} or smooth confinement\cite{kells12}.  A second, longer-term motivation of gate-defined networks involves braiding for the observation of non-Abelian statistics and quantum information applications;\cite{alicea11,clarke,vanheck12,halperin12} see, \emph{e.g.}, the setup in Fig.~\ref{fig:exp_outlook}(b).  There the keyboard of side gates should allow one to locally tune between topological and trivial regimes in a given part of the junction and hence transport Majorana zero-modes along the network.  

Various spintronics applications are also worth investigating in HgTe-based wires.  Spin qubits and spin transistors are two natural candidates that warrant further exploration\cite{datta90,koehl11,pla12}.  Finally, it would be quite interesting to perform a similar analysis of gate-defined quantum dots in HgTe quantum wells, which may inherit the remarkable features of the wires explored here.

\begin{acknowledgements}
The authors gratefully acknowledge illuminating conversations with Jim Eisenstein, David Goldhaber-Gordon, Taylor Hughes, and Torsten Karzig.  This research was supported by the Deutsche Akademie der Naturforscher Leopoldina through grant LPDS 2011-14 (J.R.); the Alfred P.\ Sloan Foundation (J.A.); the National Science Foundation through grant DMR-1055522 (J.A.) and grant DMR-1206016 (A.Y.); and the Institute for Quantum Information and Matter, an NSF Physics Frontiers Center with support of the Gordon and Betty Moore Foundation. This work is also supported in part by a grant from the Microsoft Corporation (A.Y.).
\end{acknowledgements}

\appendix
\section{Solution to the confinement problem}
\label{appendix1}

This Appendix provides details for the analytic solution of the confined wavefunctions and energies in a gate-defined HgTe wire.  As in Sec.\ \ref{subsec:band}
 we consider a Hamiltonian $H = H_{2D} + H_{\rm conf}$ as defined in Eqs.\ (\ref{general_ham}) and (\ref{Hconf}), set the parameters $\mu_{2D} = B = D = 0$ in $H_{2D}$ for simplicity, and assume the step-like confinement potential of depth $V$ illustrated in Fig.\ \ref{fig:bands}(a).  The solution proceeds by projecting onto confined states using Eq.\ (\ref{projection}) and then solving the Hamiltonian separately in regions I, II, and III labeled in Fig.\ \ref{fig:bands}(a) subject to the boundary condition that the wavefunctions are continuous.  (The wavefunctions follow from a first-order differential equation when $B = D = 0$.)  Since the Hamiltonian is block diagonal it suffices to focus on eigenstates $\Phi_{n,+}(k_x,y)$ of the upper $2\times 2$ block; eigenstates of the lower $2\times 2$ block are related by time-reversal symmetry.  
 
Consider first region II, where the solutions are described by plane waves $\propto e^{\pm ik_y y}$.   Without loss of generality we take $V > 0$ so that confined states emerge from the upper half of the Dirac cone.  The most general solution for the confined wavefunctions in region II then reads
\begin{align}
(\Phi_{n,+})_{II}
&=
b\,e^{ik_y y}\left(
\begin{array}{c}
A(k_x+ik_y)\\ -M + \sqrt{M^2+A^2(k_x^2+k_y^2)}
\end{array}
\right)\notag\\
&+
c\,e^{-ik_y y}\left(
\begin{array}{c}
A(k_x-ik_y)\\ -M + \sqrt{M^2+A^2(k_x^2+k_y^2)}
\end{array}
\right),
  \label{wf_I}
\end{align}
with corresponding energies
\begin{equation}
E_{II}=-V + \sqrt{M^2+A^2(k_x^2+k_y^2)}\,.\label{energyII}
\end{equation}
In region I, solutions are evanescent waves $\propto e^{\kappa y}$ with $\kappa>0$:
\begin{equation}
(\Phi_{n,+})_{I}
=
a\,e^{\kappa y}\left(
\begin{array}{c}
A(k_x+\kappa)\\ -M+\sigma\sqrt{M^2+A^2(k_x^2-\kappa^2)}
\end{array}
\right)\,,\label{wf_II}
\end{equation}
\begin{equation}
E_{I}=\sigma\sqrt{M^2+A^2(k_x^2-\kappa^2)}\,.
\label{EI}
\end{equation}
Here $\sigma = \pm 1$ represents the sign of the energy for a given confined state.  Similarly, in region III we obtain a solution with $\kappa\rightarrow-\kappa$,
\begin{equation}
(\Phi_{n,+})_{III}
=
d\,e^{-\kappa y}\left(
\begin{array}{c}
A(k_x-\kappa)\\ -M+\sigma\sqrt{M^2+A^2(k_x^2-\kappa^2)}
\end{array}
\right)\label{wf_III}
\end{equation}
\begin{equation}
E_{III}=E_I\,.\label{energy_III}
\end{equation}

Imposing continuity of the wavefunctions at the endpoints of region II (\emph{i.e.}, at $y = \pm W/2$) yields a set of four homogeneous equations for the four constants $a$, $b$, $c$, $d$ appearing above.  A non-trivial solution for these parameters exists provided the determinant of the corresponding matrix vanishes. This condition can be expressed as
\begin{align}
&-2\cos(k_y W) Z_+ Z_- k_y \kappa+ \sin(k_y W)\times\notag\\
&\times\left[(k_x^2-\kappa^2)Z_+^2+(k_x^2+k_y^2)Z_-^2-2k_x^2Z_+Z_-\right]=0\label{cond1}
\end{align}
with
\begin{align}
Z_+&=-M+\sqrt{M^2+A^2(k_x^2+k_y^2)}\notag\\
Z_-&=-M+\sigma\sqrt{M^2+A^2(k_x^2-\kappa^2)}.
\end{align}
As a further condition, the energies obtained in each region must of course be equal; hence
\begin{equation}
-V + \sqrt{M^2+A^2(k_x^2+k_y^2)}=\sigma\sqrt{M^2+A^2(k_x^2-\kappa^2)}\,.\label{cond2}
\end{equation}
The two conditions in Eqs.~(\ref{cond1}) and (\ref{cond2}) are sufficient to determine $k_y$ and $\kappa$, which depend both on $k_x$ and the band index $n$.  These parameters can be obtained numerically, yielding confined-band energies of the form
\begin{equation}
E_n(k_x) = -V+\sqrt{M^2+A^2[k_x^2+k_{y,n}(k_x)^2]}\,.
\end{equation}
Furthermore, by requiring normalization of the wavefunctions the constants $a, b, c, d$ can then also be determined uniquely up to an unimportant overall phase.

\section{Effective $g$-factor due to orbital effects}\label{appendix2}

Section \ref{subsec:g} discussed the orbital contribution to the gate-defined wire's effective $g$-factor, which followed from the integral in Eq.\ (\ref{analyt_int}).  Here we show how one can evaluate this integral analytically, yielding a result that is remarkably insensitive to details of the confined states.  The first important step in the calculation is to observe that for $k_x=0$ the upper and lower components of $\Phi_{n,+}$ (respectively denoted by $\phi_{n,E+}$ and $\phi_{n,H+}$) are related to one another.  Indeed, by inspecting Eq.\ (\ref{wf_I}) one can show that in region II of Fig.\ \ref{fig:bands}(a) (\emph{i.e.}, for $|y|< W/2$) the wavefunction components satisfy
\begin{align}
\partial_y\phi_{n,E+}(0,y)&=\frac{-Ak_y^2}{-M+\sqrt{M^2+A^2k_y^2}}\phi_{n,H+}(0,y)\,,\label{wf_relation1}\\
\partial_y\phi_{n,H+}(0,y)&=\frac{-M+\sqrt{M^2+A^2k_y^2}}{A}\phi_{n,E+}(0,y)\,.
\end{align}
Similar relations hold in regions I and III (\emph{i.e.}, for $|y|\geq W/2$):
\begin{align}
\partial_y\phi_{n,E+}(0,y)&=\frac{A\kappa^2}{-M+\sigma\sqrt{M^2-A^2\kappa^2}}\phi_{n,H+}(0,y)\,,\\
\partial_y\phi_{n,H+}(0,y)&=\frac{-M+\sigma\sqrt{M^2-A^2\kappa^2}}{A}\phi_{n,E+}(0,y)\,.\label{wf_relation4}
\end{align}

Consider next the normalization condition at $k_x = 0$,
\begin{equation}
  1=\int_{-\infty}^{\infty}dy\left[\phi_{n,E+}(0,y)^2+\phi_{n,H+}(0,y)^2\right].
\end{equation}
The right-hand side can in fact be recast into a form very similar to the integral in Eq.\ (\ref{analyt_int}) that we are trying to evaluate.  Specifically, upon inserting a trivial factor $\partial_y y$ (which equals unity) under the integral and then integrating parts, we obtain
\begin{eqnarray}
   1 &=& -2\int_{-\infty}^{\infty}dy[\phi_{n,E+}(0,y)y\partial_y\phi_{n,E+}(0,y)
   \nonumber \\
   &+&\phi_{n,H+}(0,y)y\partial_y\phi_{n,H+}(0,y)]\,.
\end{eqnarray}
Finally, we break the right side up into separate integrals over regions  I, II, and III and employ the relations from Eqs.~(\ref{wf_relation1})-(\ref{wf_relation4}) to eliminate the derivatives. After some algebra the three parts of integration can be reconciled, yielding
\begin{equation}
1=\frac{4M}{A}\int_{-\infty}^{\infty} dy \phi_{n,E+}(0,y) y\phi_{n,H+}(0,y)\,,
\end{equation}
which immediately proves Eq.~(\ref{analyt_int}).

\section{Oscillations arising from non-perturbative effects of Rashba coupling}\label{appendix3}

In this final Appendix we provide a detailed account of the effects of Rashba spin-orbit coupling in the limit where a `large' perpendicular electric field impinges on the quantum well.  (The meaning of `large' is clarified in Sec.\ \ref{subsec:eso}.)  More precisely, our goal is to understand the dramatic oscillations in the gate-defined wire's effective $g$-factor and spin-orbit energy induced by varying the voltage drop $U$ generated by the field (recall Fig.\ \ref{fig:g_eso}).  Ultimately these features reflect strong modifications of the confined-state wavefunctions by the perpendicular electric field, and it is therefore essential that one treats Rashba coupling \emph{non-perturbatively} here.  Since this poses a nontrivial analytic task we will study a simplified model that facilitates progress yet still captures the essential physics.  

First, as in Appendix \ref{appendix1} we neglect terms quadratic in momenta in the 2D quantum well Hamiltonian and assume a step-like confinement potential $V(y)=-V\Theta(W/2-|y|)$ to define the HgTe wire.  Second, we restrict our considerations to deep confinement potentials and high subbands, where the oscillations are most pronounced as Fig.\ \ref{fig:g_eso} illustrates. In other words, we assume $V \gg M$ so that the Dirac cone inside the confined region II in Fig.\ \ref{fig:bands}(a) is, roughly speaking, strongly shifted relative the cones in the surrounding regions I and III.  The Dirac mass $M$ in region II then negligibly impacts the confined states and can be safely ignored.  Third, the Rashba coupling in regions I and III does not significantly influence the oscillations we aim to describe, so for simplicity we will retain Rashba coupling only within region II (which rather naturally provides the dominant effect on the confined states).  

The full Hamiltonian we treat is then $H = H_{2D} + H_{\rm conf} + H_R$, where the terms on the right are defined in Eqs.\ (\ref{general_ham}), (\ref{Hconf}), and (\ref{ham_rashba}); given the assumptions above we set $B = D = 0$, $M \rightarrow M\Theta(|y|-W/2)$, and replace $R\rightarrow R\Theta(W/2-|y|)$ in the Rashba term.  As a final simplification we content ourselves with solving $H$ above for confined wavefunctions with momentum $k_x = 0$.  This suffices for capturing directly the oscillations in the effective $g$-factor but not the effective Rashba energy scale characterizing the gate-defined wire, for which one would also need information about finite-$k_x$ states.  Nevertheless, indirect arguments for Rashba oscillations can be made as described below.

In the following, we proceed as in Appendix~\ref{appendix1} and discuss the confined wavefunctions and energies at $k_x = 0$ by treating regions I, II, and III of Fig.\ \ref{fig:bands}(a) separately and then imposing proper boundary conditions.  Note that the form of $H$ guarantees that each component of the $k_x = 0$ wavefunctions has definite parity under $y\rightarrow -y$.  More precisely, one can show that
\begin{align}
&[\phi_{E+}(y),\phi_{H+}(y),\phi_{E-}(y),\phi_{H-}(y)]\notag\\
=\lambda &[\phi_{E+}(-y),-\phi_{H+}(-y),-\phi_{E-}(-y),\phi_{H-}(-y)]\,,\label{symm_relation}
\end{align}
where solutions with $\lambda = +1$ and $-1$ correspond to Kramer's pairs.  (For notational simplicity, here and below we suppress the band index $n$ on the wavefunctions; furthermore, all wavefunctions implicitly refer to $k_x=0$.)  We therefore need only solve explicitly for the wavefunctions in regions II and III since the form in region I follows from Eq.\ (\ref{symm_relation}).  

We begin with region II, where the $k_x = 0$ wavefunctions are eigenstates of 
\begin{equation}
H_{II}=
\left(\begin{array}{cccc}
-V                  & A\partial_y & iR\partial_y &0\\
-A\partial_y & -V                 & 0                   &0\\
iR\partial_y & 0                   & -V                 &A\partial_y\\
0                   & 0                   & -A\partial_y &-V
\end{array}\right)\,.
\end{equation}
Diagonalizing this matrix using a plane-wave ansatz $\propto e^{\pm ik_y y}$ yields the four energies
\begin{align}
E_{II,1}&=|k_y|\frac{R+\sqrt{4A^2+R^2}}{2}-V\notag\\
E_{II,2}&=|k_y|\frac{-R+\sqrt{4A^2+R^2}}{2}-V\notag\\
E_{II,3}&=-E_{II,1}\,,\,E_{II,4}=-E_{II,2}\,.
\end{align}
Let us assume that the confinement potential $V$ is positive so that the confined states emerge from the upper half of the Dirac cone.  In this case only the upper branches $E_{II,1}$ and $E_{II,2}$ are relevant so we discard the others.  Most crucially, due to Rashba coupling there are now two distinct sets of momenta $\pm k_{y1}$ and $\pm k_{y2}$, which are related via
\begin{equation}
  k_{y2}=k_{y1}\frac{R+\sqrt{4A^2+R^2}}{-R+\sqrt{4A^2+R^2}},
  \label{relation_k1_k2}
\end{equation} 
that yield the same energy.  This fact reflects the usual Rashba splitting of bands [illustrated in Fig.\ \ref{fig:cone_rashba}(b) for $M \neq 0$] and is intimately related to the appearance of oscillations.  

Consider next the corresponding confined wavefunctions, which generically consist of superpositions of plane waves with all four $k_y$ values above (since momentum is not conserved along $y$).  Using Eq.~(\ref{symm_relation}) with $\lambda = +1$ along with Eq.\ (\ref{relation_k1_k2}), wavefunctions in region II take the form
\begin{widetext}
\begin{equation}
\left(\begin{array}{c}
\phi_{E+}^+(y)\\ \phi_{H+}^+(y)\\ \phi_{E-}^+(y)\\ \phi_{H-}^+(y)
\end{array}\right)_{II}
=a_+
\left(\begin{array}{c}
(R+\sqrt{4A^2+R^2})\cos(k_{y1}y)\\ 2A\sin(k_{y1}y)\\-i(R+\sqrt{4A^2+R^2})\sin(k_{y1}y) \\ 2iA\cos(k_{y1}y)
\end{array}\right)
+b_+
\left(\begin{array}{c}
(-R+\sqrt{4A^2+R^2})\cos(k_{y2}y)\\ 2A\sin(k_{y2}y)\\i(-R+\sqrt{4A^2+R^2})\sin(k_{y2}y) \\ -2iA\cos(k_{y2}y)
\end{array}\right)\,,\label{wf_II_rashba}
\end{equation} 
\end{widetext}
where $a_+$ and $b_+$ are constants and the added superscript on the wavefunction components indicates that $\lambda = +1$.  [For the Kramer's partner $\phi^-_{E/H\pm}$ with $\lambda = -1$ one simply swaps cosines and sines, \emph{i.e.},  $\sin(\cdots)\rightarrow -i\cos(\cdots)$, $\cos(\cdots)\rightarrow i\sin(\cdots)$.]  

In region III the $k_x = 0$ wavefunctions are eigenstates of 
\begin{equation}
H_{III}=
\left(\begin{array}{cc}
\tilde{H}_{III}&0\\
0&\tilde{H}_{III}
\end{array}\right)
\,\text{with}\,\,
\tilde{H}_{III}=
\left(\begin{array}{cc}
M&A\partial_y\\
-A\partial_y&-M
\end{array}\right)\,.
\end{equation}
Since there is no Rashba coupling here the solutions can be immediately read off from Eqs.~(\ref{wf_III}) and (\ref{energy_III}):
\begin{equation}
\left(\begin{array}{c}
\phi_{E+}^\lambda(y)\\ \phi_{H+}^\lambda(y)\\ \phi_{E-}^\lambda(y)\\ \phi_{H-}^\lambda(y)
\end{array}\right)_{III}
=
\left(\begin{array}{c}
-c_{\lambda}A\kappa\\ c_{\lambda}(-M+\sigma\sqrt{M^2-A^2\kappa^2})\\-d_{\lambda}A\kappa \\ d_{\lambda}(-M+\sigma\sqrt{M^2-A^2\kappa^2})
\end{array}\right)e^{-\kappa y}\,,
\end{equation}
with energies
\begin{equation}
E_{III}=\sigma\sqrt{M^2-A^2\kappa^2}\,.\label{energy_III2}
\end{equation}
As before $\sigma = \pm 1$ follows from the sign of the energy, the superscript $\lambda$ in the wavefunctions labels Kramer's partners, and $c_{\lambda}$ and $d_{\lambda}$ are constants.  
 
The wavefunctions must be continuous at $y = W/2$, which leads to a set of four homogeneous equations for $a_\lambda$, $b_\lambda$, $c_\lambda$, and $d_\lambda$.  Again, for a non-trivial solution to exist the corresponding matrix must have zero determinant.  After some algebra, this condition can be written as
\begin{equation}
\tan\left[\frac{(k_{y1}+k_{y2})W}{2}\right]=\sigma\frac{\kappa}{2}\sqrt{\frac{4A^2+R^2}{M^2-A^2\kappa^2}}\,,\label{det_zero}
\end{equation}
which holds for either $\lambda=+1$ or $-1$. The energies in regions II and III must also match, yielding
\begin{equation}
\frac{k_{y1}}{2}(R+\sqrt{4A^2+R^2})-V=\sigma\sqrt{M^2-A^2\kappa^2}\,.\label{energy_match}
\end{equation}
Equations (\ref{relation_k1_k2}), (\ref{det_zero}) and (\ref{energy_match}) are sufficient to determine $k_{y1}$, $k_{y2}$ and $\kappa$. According to Eq.~(\ref{det_zero}) solutions for different bands correspond to the intersection points of the different branches of the tangent with the $\kappa$-dependent right-hand side.  Hence, by changing the Rashba coupling $R \propto U$, the sum $k_{y1}+k_{y2}$ for band $n$ can only vary in the interval $k_{y1}+k_{y2}\in[2(n-1)\pi/W,(2n-1)\pi/W]$ (for positive energies) or $k_{y1}+k_{y2}\in[(2n-1)\pi/W,2n\pi/W]$ (for negative energies).   The actual variation with $U$ is typically much smaller, and for the high bands of interest here one can approximate the sum as a constant, $k_{y1}+k_{y2}\approx 2k_0$. On the other hand, with Eqs.~(\ref{relation_k1_k2}), (\ref{energy_III2}) and (\ref{energy_match}), one can show that the difference yields
\begin{equation}
k_{y1}-k_{y2}=-\frac{(V+E)R}{A^2}\approx -\frac{VR}{A^2}\,.
\end{equation}
In the last step we used $E\ll V$ as appropriate for the deep confinement potentials we are considering. The two momenta are therefore well-approximated by
\begin{equation}
k_{y1}\approx k_0-\frac{V}{2A^2}R\,,\,\,~~~~k_{y2}\approx k_0+\frac{V}{2A^2}R\,.\label{approx_k1_k2}
\end{equation}
\begin{figure}[t]
\centering
\includegraphics[scale=0.32]{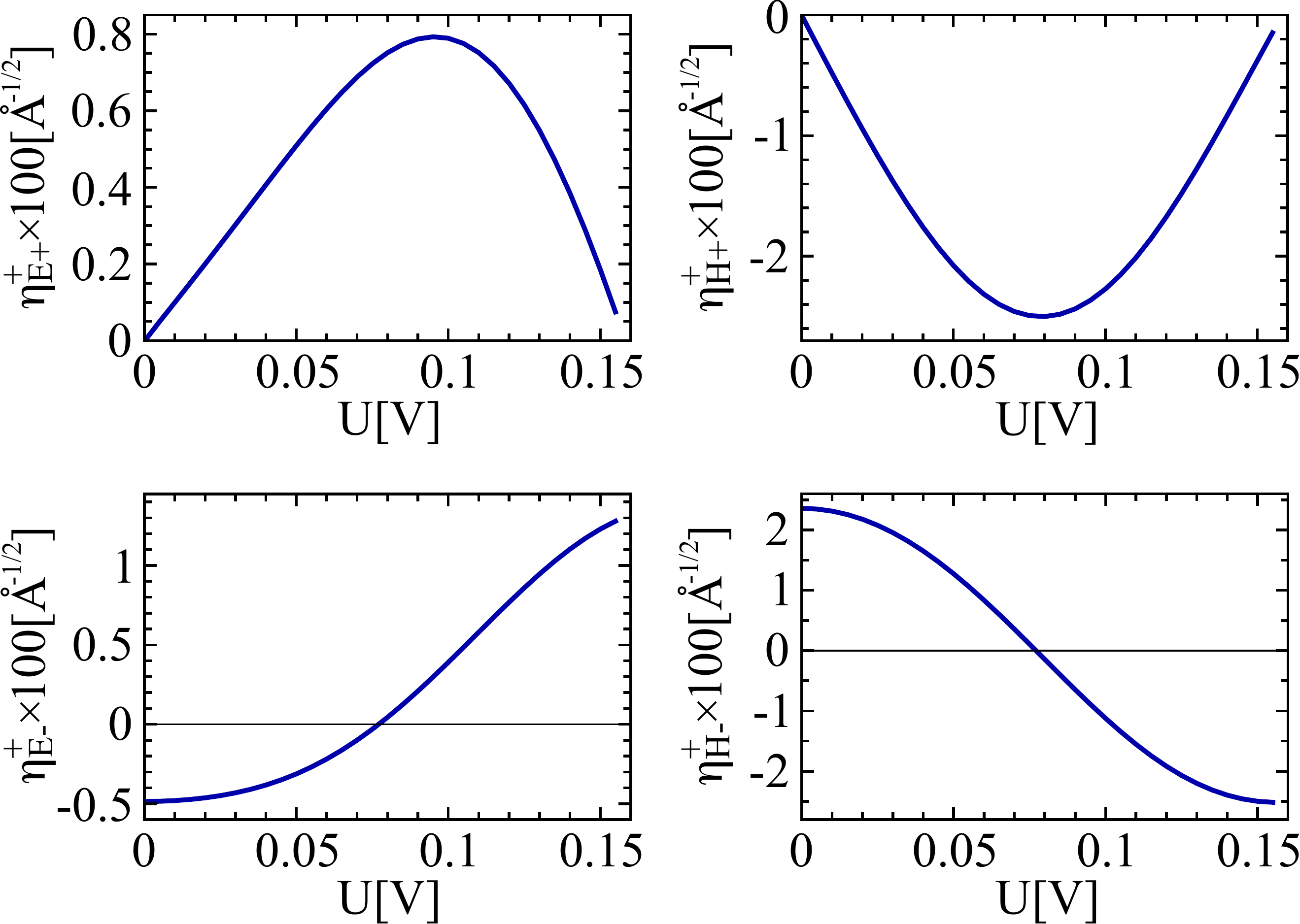}
\caption{Oscillating behavior for components of one of the $k_x = 0$ wavefunctions evaluated at $y = W/2$, corresponding to the edge of the gate-defined wire.  In the plots $U$ is the voltage drop induced by a perpendicular electric field, $\eta_{E/H+}^+\equiv\phi_{E/H+}^+(y=W/2)$, and $\eta_{E/H-}^+\equiv i \phi_{E/H-}^+(y=W/2)$.  The data were obtained numerically by considering the $n = 5$ band in a HgTe system with $d=70$\AA, $W = 800$\AA~and $V = 0.05$eV.  }
\label{fig:osc}
\end{figure}

Next, we consider the behavior of the wavefunctions at $y=W/2$ and define $\eta_{E/H+}^\lambda\equiv\phi_{E/H+}^\lambda(y=W/2)$ and $\eta_{E/H-}^\lambda\equiv i \phi_{E/H-}^\lambda(y=W/2)$ (the factor of $i$ is inserted so that all $\eta$'s can be chosen real, which we henceforth assume is the case). By inspecting Eq.~(\ref{wf_II_rashba}) and its time-reversed partner one sees that the $\eta$'s for a given Kramer's pair are related via
\begin{equation}
  \eta_{E/H+}^-=\eta_{E/H-}^+\,,\qquad\eta_{E/H-}^-=-\eta_{E/H+}^+\,.\label{eta_rel}
\end{equation}
Equation (\ref{wf_II_rashba}) also implies that, schematically, for $\lambda=+1$ these components vary with $k_{y1,2}$ according to
\begin{align}
\eta_{E+/H-}^+ &\sim c_1\cos\left(\frac{k_{y1}W}{2}\right)+c_2\cos\left(\frac{k_{y2}W}{2}\right)\,,\notag\\
\eta_{H+/E-}^+ &\sim c_3\sin\left(\frac{k_{y1}W}{2}\right)+c_4\sin\left(\frac{k_{y2}W}{2}\right)\,.\label{oscillations}
\end{align}
It then follows from Eq.~(\ref{approx_k1_k2}) that the arguments of the sines and cosines are given by $\frac{k_0W}{2}\pm\frac{VW}{4A^2}R$, leading to oscillations in the wavefunction components as a function of $R$ with the same period $\Delta R=\frac{8\pi A^2}{VW}$. Note that $\eta_{E/H+}^+$ and $\eta_{E/H-}^+$ are out of phase with one another.  This important property is evident in Fig.~\ref{fig:osc}, which illustrates these oscillations for the $n = 5$ band in a system with $d=70$\AA, $W = 800$\AA~and $V = 0.05$eV; the data were obtained by numerically solving Eqs.~(\ref{relation_k1_k2}), (\ref{det_zero}) and (\ref{energy_match}).  Crucially, because the wavefunctions are continuous at $y = W/2$, the amplitudes of the exponential tails in region III (and by symmetry region I) follow the same oscillations.  Thus Rashba coupling can, quite remarkably, be tuned such that for certain wavefunction components the exponential tail is completely suppressed. 

Let us now use these results to capture oscillations in the gate-defined wire's effective $g$-factor.  We focus on the orbital contribution to the $g$-factor since as we saw in Sec.\ \ref{subsec:g} this generally dominates over the conventional Zeeman terms.  Of interest then is the energy splitting of $k_x = 0$ Kramer's pairs in a given band in response to the orbital part of the field.  Upon sending $\partial_x \rightarrow \partial_x-i ey \mathcal{B}/\hbar$ in the 2D quantum well Hamiltonian (again with $B = D = 0$), first-order perturbation theory gives a splitting 
\begin{align}
  \Delta E_{\rm orb} &= \frac{2eA\mathcal{B}}{\hbar}\int  dy\,y\left[\phi_{E+}^{+*}(y) \phi_{H+}^+(y)+\phi_{H+}^{+*}(y) \phi_{E+}^+(y)\right.\notag\\
&-\left.\phi_{E-}^{+*}(y) \phi_{H-}^+(y)-\phi_{H-}^{+*}(y) \phi_{E-}^+(y)\right]\,.\label{int_orbital2}
\end{align}
Note that this expression only involves wavefunctions with the same $\lambda$---off-diagonal matrix elements vanish for symmetry reasons.  Orbital effects from the Rashba term are also neglected for simplicity since they provide a small correction.  

Contributions from `small' $y$ in the integral in Eq.\ (\ref{int_orbital2}) are suppressed both by the factor $y$ appearing in the integrand, as well as nodes in the wavefunctions present in the confined region (recall that we are treating high subbands).  Thus the integral is dominated by values $|y|\gtrsim W/2$.  One can then restrict the range of integration to this regime and express Eq.~(\ref{int_orbital2}) in terms of $\eta$'s.  We then obtain, using symmetry properties of the wavefunctions under $y\rightarrow-y$,
\begin{equation}
  \Delta E_{\rm orb}\approx\frac{8eA\mathcal{B}}{\hbar}(\eta_{E+}^+\eta_{H+}^+-\eta_{E-}^+\eta_{H-}^+)\3\int_{W/2}^{\infty}\2\3dy\,ye^{-2\kappa \left(y-\frac{W}{2}\right)}\,.
  \label{g-factor_osc}
\end{equation}

The appearance of $g$-factor oscillations is now manifest. When $\eta_{E/H+}^+$ are both large while $\eta_{E/H-}^+$ are both small (\emph{e.g.}, at $U\approx0.08$V in Fig.~\ref{fig:osc}) the former wavefunction components extend appreciably in space and hence produce a large contribution to the effective $g$-factor.  The $\eta_{E/H-}^+$ components, by contrast, are sharply suppressed away from the gate-defined wire and contribute negligibly.  Similar results hold in the opposite limit where $\eta_{E/H+}^+$ are small and $\eta_{E/H-}^+$ are large (\emph{e.g.}, at $U = 0$ in Fig.\ \ref{fig:osc}), though the sign of the splitting reverses.  If, on the other hand, all components are of similar magnitude a cancellation results and the orbital contribution to the $g$-factor vanishes.  Hence, the products of $\eta$'s in Eq.~(\ref{g-factor_osc}) effectively double the frequency of oscillation compared to that of the individual wavefunction components. Setting $R=FeU$, the period of $g$-factor oscillations in terms of the voltage $U$ is therefore given by
\begin{equation}
\Delta U=\frac{4\pi A^2}{eFVW}\,.\label{period}
\end{equation}
Note that the $1/V$-dependence of the periodicity is in accordance with Fig.~\ref{fig:g_eso}. As another check, inserting $V=0.05$eV and the parameters used in Fig.\ \ref{fig:g_eso}, one obtains a period $\Delta U\approx0.19$V that is comparable to that seen in our more accurate numerics. Deviations arise mainly from the $B$ and $D$ terms in the quantum well Hamiltonian which have been neglected here for simplicity.

Consider next the effective Rashba spin-orbit coupling induced in the gate-defined wire.  Our analysis has so far focused on $k_x = 0$, and extending the results to finite $k_x$ is nontrivial.  
Nevertheless, terms in the Hamiltonian involving $\partial_x$, which generate Rashba-induced energy corrections $\propto k_x$ as $k_x \rightarrow 0$, can be treated relatively easily in the limit of small but finite $k_x$.  These corrections can be extracted from the $k_x = 0$ wavefunctions already determined.  (Finite-$k_x$ corrections to the wavefunctions contribute only at higher-order in $k_x$.)  Two such terms exist: one arising from the $A$ term in the 2D quantum well Hamiltonian [Eq.\ (\ref{general_ham})], the other appearing directly in the Rashba Hamiltonian [Eq.\ (\ref{ham_rashba})].  In the non-perturbative Rashba coupling limit of interest, the splitting arising from the latter is limited by the Rashba coefficient $R$ while that generated by the former is limited by $A$.  Thus provided $R \lesssim A$, which roughly corresponds to the voltage regime considered here, the $A$ term generally dominates.  

Focusing on this contribution, the Rashba splitting for a given confined subband reads
\begin{align}
  \Delta E_{\rm R}&= 2Ak_x\int dy\,\left[\phi_{E+}^{+*}(y) \phi_{H+}^-(y)+\phi_{H+}^{+*}(y) \phi_{E+}^-(y)\right.\notag\\
&-\left.\phi_{E-}^{+*}(y) \phi_{H-}^-(y)-\phi_{H-}^{+*}(y) \phi_{E-}^-(y)\right]\,.\label{int_rashba2}
\end{align}
Note that in contrast to Eq.~(\ref{int_orbital2}) this integral involves coupling of Kramer's partners with opposite $\lambda$.  Since the integrand is highly oscillatory in the confined region due to the high subbands considered here, it is again a good approximation to evaluate the integral only in the region $|y|\gtrsim W/2$. With the aid of Eq.\ (\ref{eta_rel}), one can then formulate Eq.\ (\ref{int_rashba2}) in terms of $\eta_{E/H\pm}^+$, yielding
\begin{equation}
  \Delta E_{\rm R}\approx8Ak_x(\eta_{E+}^+\eta_{H-}^++\eta_{E-}^+\eta_{H+}^+)\int_{W/2}^{\infty}dy\,e^{-2\kappa \left(y-\frac{W}{2}\right)}\,.
  \label{rashba_osc}
\end{equation}
The above expression depends implicitly on $R$ through the $\eta$ terms, and vanishes linearly at small $R$ as one would expect from perturbation theory.  To see this consider the products $\eta_{E+}^+\eta_{H-}^+$ and $\eta_{E-}^+\eta_{H+}^+$.  According to Fig.~\ref{fig:osc}, at small $R$ one of the $\eta$'s in each product is roughly constant while the other increases linearly from zero with $R$; thus $\Delta E_{\rm R} \sim R$ as $R\rightarrow 0$.  

More interestingly, Eq.\ (\ref{rashba_osc}) captures the oscillations in the effective Rashba coupling for the confined subbands at larger $R$.  The physics is closely related to our discussion of the $g$-factor below Eq.\ (\ref{g-factor_osc}); in particular the oscillations again derive from electric-field-induced modulations of the confined wavefunctions' exponential tails.  Let us simply highlight two noteworthy points here.  First, the oscillation period in $U$ is the same as for the $g$-factor---see Eq.\ (\ref{period}).  Second, the Rashba and $g$-factor oscillations are \emph{out of phase}, which can be seen by comparing Eqs.\ (\ref{g-factor_osc}) and (\ref{rashba_osc}) and inspecting Fig.\ \ref{fig:osc}.  Whenever either $\eta^+_{E/H+}$ or $\eta^+_{E/H-}$ becomes zero, for example, the Rashba splitting vanishes while the $g$-factor attains a large magnitude.  This somewhat crude treatment of Rashba oscillations recovers the essential features seen in our more reliable numerics in Fig.\ \ref{fig:g_eso}.

\bibliographystyle{prsty}

\end{document}